\documentclass[12pt]{article}
\usepackage{epsfig, epsf,subfigure, color}
\usepackage{pstricks, pst-node, psfrag}
\usepackage{amssymb,amsmath}
\usepackage{verbatim,enumerate}
\usepackage{rotating, lscape}
\usepackage{setspace}
\usepackage{bbm}
\usepackage{natbib}
\usepackage{amsthm}
\usepackage{dsfont}
\usepackage{url}
\usepackage{multirow}
\usepackage{amsmath}
\usepackage{amssymb}
\usepackage{appendix}
\usepackage{tabularx,booktabs}
\usepackage{amsmath}

\newcommand{\Rlogo}
{\protect\includegraphics[height=1.8ex,keepaspectratio]{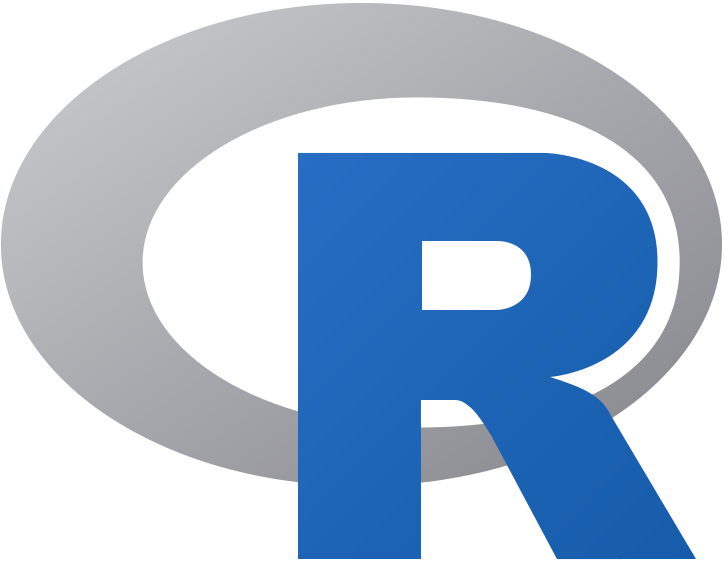}}
\graphicspath{{plots/}}
\usepackage{graphicx}
\usepackage{epstopdf}
\usepackage{tikz-cd}
\DeclareGraphicsExtensions{.pdf}
\newcommand{\pkg}[1]{{\normalfont\fontseries{b}\selectfont #1}}

\usepackage[shortlabels]{enumitem}

\DeclareMathOperator*{\argmin}{arg\,min}
\usepackage{enumitem}
\usepackage{float}
\usepackage[colorlinks=true,citecolor=blue]{hyperref}
\usepackage{amsmath,amsfonts,amsthm,bm}
\setlist[enumerate]{wide=0pt, widest=99,leftmargin=\parindent, labelsep=*}
\begin{document}

\thispagestyle{empty} \baselineskip=28pt \vskip 5mm
	\begin{center} {\Large{\bf A semi-parametric estimation method for  quantile coherence  with an application to bivariate financial time series clustering}}
	\end{center}
	
	\baselineskip=12pt \vskip 5mm
	
	\begin{center}\large
		Cristian F. Jim\'enez-Var\'on\footnote[1]{
			\baselineskip=10pt CEMSE Division, Statistics Program, King Abdullah University of Science and Technology, Thuwal 23955-6900, Saudi Arabia. E-mail: cristian.jimenezvaron@kaust.edu.sa;  ying.sun@kaust.edu.sa}$^{*}$, Ying Sun$^{1}$, Ta-Hsin Li~\footnote[2]{
			\baselineskip=10pt IBM Watson Research Center, NY, United States of America. E-mail: thl@us.ibm.com }\end{center}
	
	\baselineskip=16pt \vskip 1mm \centerline{\today} \vskip 8mm

	\begin{center}
		{\large{\bf Abstract}}
	\end{center}
	\baselineskip=17pt
 In multivariate time series analysis, spectral coherence measures the linear dependency between two time series at different frequencies. However, real data applications often exhibit nonlinear dependency in the frequency domain. Conventional coherence analysis fails to capture such dependency. The quantile coherence, on the other hand, characterizes nonlinear dependency by defining the coherence at a set of quantile levels based on trigonometric quantile regression. This paper introduces a new estimation technique for quantile coherence. The proposed method is semi-parametric, which uses the parametric form of the spectrum of a vector autoregressive (VAR) model to approximate the quantile coherence, combined with nonparametric smoothing across quantiles. At a given quantile level, we compute the quantile autocovariance function (QACF) by performing the Fourier inverse transform of the quantile periodograms. Subsequently, we utilize the multivariate Durbin-Levinson algorithm to estimate the VAR parameters and derive the estimate of the quantile coherence. Finally, we smooth the preliminary estimate of quantile coherence across quantiles using a nonparametric smoother. Numerical results show that the proposed estimation method outperforms nonparametric methods.  We show that quantile coherence-based bivariate time series clustering has advantages over the ordinary VAR coherence. For applications, the identified clusters of financial stocks by quantile coherence with a market benchmark are shown to have an intriguing and more informative structure of diversified investment portfolios that may be used by investors to make better decisions.

	\begin{doublespace}
		
		\par\vfill\noindent
		{\bf Some key words}: Clustering, Financial stocks,  Financial time series, Quantile coherence, Quantile spectral analysis, VAR approximation.
	\par\medskip\noindent
		{\bf Short title}: Semi-parametric quantile coherence estimation
	\end{doublespace}
	
	\clearpage\pagebreak\newpage \pagenumbering{arabic}
	\baselineskip=26.5pt
	
\section{Introduction}\label{sec:intro}

For univariate time series, motivated by the least-squares interpretation of the ordinary periodogram (OP),  \cite{Li2012} proposed the quantile periodogram (QPER) based on trigonometric quantile regression. The QPER, like the ordinary periodogram, has an asymptotic exponential distribution where the mean function, known as the quantile spectrum, is a scaled version of the ordinary spectrum of the level-crossing process. In  \cite{Li2012} the QPER is shown to be resistant to nonlinear data distortion in the sense that any nonlinear memoryless transformation only affects the scaling constant in the asymptotic distribution in the given quantile. \cite{memoryless} shows that OP is not resistant to such nonlinear data transformations as the power spectrum is distorted.  

The QPER can be extended to multivariate time series, and in particular, the quantile coherence quantifies the nonlinear dependency across time series by defining coherence as a function of quantile levels as well as frequencies. The quantile coherence has an advantage over its ordinary coherence counterpart as its asymptotic distribution relies on the level-crossing cross-spectrum, which, like QPER, is resistant to nonlinear data distortion due to the robustness of quantile regression \citep{Li2014}. Even though quantile coherence is a more effective tool, how to estimate it remains an interesting question.

Periodogram smoothing is the first that comes to mind. In time series analysis, many methods exist to smooth periodograms for univariate time series. For instance, \cite{Shumway2017} presented several nonparametric periodogram smoothing approaches that can be used across frequencies, such as moving-average smoothing. The optimally smoothed spline (OSS) estimator by \cite{Wahba1980} selects the smoothing parameter by minimizing the expected integrated mean square error. Some other common spectral estimations are based on the likelihood. For instance, \cite{Capon1983} used a high-resolution estimation method, known as the maximum-likelihood method (MLM). MLM filters the time series to produce the minimum-variance unbiased estimator of the spectrum. For bivariate time series data, \cite{Pawitan1994} estimated the spectrum by means of the penalized Whittle likelihood. In terms of parametric estimators, \cite{burg1975} presented the maximum entropy type of estimators, which in the univariate case results in the autoregressive (AR) spectral approximation. \cite{Dette2015}  demonstrated that a window smoother of the quantile periodogram may consistently estimate the quantile spectrum across frequencies. \cite{Chen2020} employed a semi-parametric estimation of the quantile spectrum by using the approximation capability of the AR model. A nonparametric smoother is proposed to smooth both quantile levels and AR coefficients in the partial autocorrelation function (PACF) domain. \cite{QC} introduced a nonparametric quantile coherency estimator based on the level-crossing process; in addition, to achieve consistency of the estimator, they proposed smoothing the proposed copula cross-periodogram across frequencies based on the results for the univariate case, initially introduced by \cite{Kley2016} (Proposition 3.4).

In this paper, we introduce an alternative technique for estimating quantile coherence. Our proposed method is semi-parametric. It combines the parametric version of the (VAR) spectrum with nonparametric smoothing across quantile levels to estimate the quantile coherence. Our estimator is based on the assumption that true quantile coherence, as a bivariate function of frequencies and quantiles, exhibits smoothness across quantile levels.

A key distinction between our approach and the method introduced by \cite{QC} lies in the nature of the estimated quantile coherence. Our estimator considers both frequencies and quantile levels as bivariate inputs, encompassing a range of quantile values in the interval (0,1) rather than a fixed level.  Furthermore, the estimator proposed by \cite{QC} employs kernel smoothing across frequencies, requiring the precise selection of a bandwidth parameter and the type of kernel for consistency. In contrast, our semi-parametric approach utilizes the VAR representation of the quantile spectra at each quantile level. The order of the VAR can be determined using an automatic selection method such as AIC, which simplifies the process compared to optimal bandwidth selection.

The concept of employing a parametric autoregressive (AR) model for spectrum estimation has been advocated by researchers such as \cite{Akaike1969-su, Parzen1983, Newton1981}. Additionally, the maximum entropy framework has been used to characterize autoregressive spectral densities as models for the spectral density of a stationary time series \citep{burg1975, Choi1993, Parzen1982}. In the univariate case, the spectral density function of a stationary process can be approximated by the spectral density function of an autoregressive (AR)  process \cite{Shumway2017} (Property 4.7, see Appendix C.6 for more details). This property also holds for other time series representations, such as the moving average (MA) process or the autoregressive moving average (ARMA) process.
For multivariate time series, \cite{Wiener1957I} (Theorem 7.13) demonstrates that a given non-negative Hermitian matrix-valued function can be factorized into the product of two functions defined over the complex plane. \cite{Wiener1958II} (Section 2) introduces the vector autoregressive (VAR) as a means of determining the matrix coefficients of the generating functions in the factorization representation. While we have not conducted a thorough analysis of the theoretical properties of our proposed estimator, empirical studies indicate its effectiveness in practical applications. We acknowledge that further theoretical development is a potential avenue for future research.

Our proposed parametric approach for estimating quantile coherence is a three-step process. Firstly, we derive the quantile autocovariance function (QACF) from the quantile periodograms at fixed quantile levels. Secondly, we compute the VAR parameters based on the QACF using the Durbin-Levinson algorithm. The order of the VAR approximation for all quantile levels is automatically selected using the AIC criterion \citep{Akaike1974}. Thirdly, we obtain the initial estimate of quantile coherence using the VAR models. Finally, we employ a nonparametric smoother to refine the initial estimate of quantile coherence along quantile levels for each frequency. Under the assumption that the degree of smoothness across quantiles does not change dramatically at different frequencies, we introduce a cross-validation criterion to select the common tuning parameter for all frequencies. To evaluate the performance of our estimation method, we provide some empirical evidence through a simulation study.

Spectral features, including spectral coherence, have been used as input for many applications in time series classification and clustering \citep{caro,Chen2020}.  \cite{Maadooliat2018} applied their proposed spectral density estimation methods for brain signal clustering. \cite{caro}  developed a coherence-based hierarchical clustering method with application to brain connectivity.  The quantile frequency analysis (QFA) \citep[see, e.g.,][]{Li2020,Li2021} uses a two-dimensional function of the quantile spectral estimate by varying the quantile level as well as the trigonometric frequency parameter. The QFA method has been used in two ways in classification. The first method treats quantile periodograms as images that can be directly fed into a deep-learning image classifier like convolutional neural networks (CNN). The second method employs dimension-reduction techniques and feeds the resulting features into a general-purpose classifier such as the support vector machine \citep{Hastie2017}. \cite{Chen2020} used the former approach to classify earthquake waves. \cite{Li2020} used the latter approach to classify real-world ultrasound signals for nondestructive evaluation of the structural integrity of aircraft panels. 

Financial time series clustering was examined by \cite{Bishnoi2018} using a quantile periodogram approach with stock price data from various sectors. In the realm of risk analysis, \cite{BN2022} introduced the concept of quantile spectral beta (QSB) to represent risk, leveraging the quantile cross-spectral densities proposed by \cite{QC} to capture tail and spectral risks at a specific quantile level.

In the application section, we investigate the problem of clustering 52 stocks based on their quantile coherence with the S\&P 500 (SPX) index at specific quantile regions. For each stock, the daily log returns are coupled with the corresponding values from the S\&P 500 closing prices to a 2D vector time series from which the quantile coherence is computed. The quantile coherence takes into account both frequencies and quantile levels and serves as an input for hierarchical clustering. To evaluate the clusters produced using quantile coherence, we compare them with the clusters generated using the ordinary coherence equivalent. Furthermore, we analyze the behavior of the clusters obtained through the standard time domain approach, which uses the beta coefficient in the Capital Asset Pricing Model (CAPM). Our analyses reveal that quantile coherence not only provides additional valuable information but also tends to produce more meaningful clusters.

The rest of the paper is organized as follows. In Section \ref{sec:method}, we introduce the quantile coherence and the proposed estimation procedure. The simulation study is presented in Section \ref{sec:simualtion}, and the application to financial time series clustering is described in Section \ref{sec:application}. We conclude and discuss the paper in Section \ref{sec:discussion}.

\section{Methodology}\label{sec:method}

In this section, we first introduce the quantile spectrum, the quantile periodogram, and the quantile coherence in Section \ref{method:quantile spectrum and quantile coherence}. In Section \ref{methods: YW approach}, we present the VAR spectral model and its estimation. The VAR representation of the estimated quantile spectral matrix is presented in Section \ref{methods: VAR approx}.  Finally, in Section \ref{methods: proposed smoothing}, we present the proposed smoothing procedure for the VAR spectrum.

\subsection{Quantile spectrum and quantile coherence}\label{method:quantile spectrum and quantile coherence}

Assume ${Y_{t,j}}~(j=1,\ldots,k)$ are $k$ jointly stationary random processes. Let $F_j(u)$ denote the CDF of ${Y_{t,j}}$ which is a continuous function with the derivative $\dot{F_j}(u)>0$. Let $\lambda_{j,\alpha}$ denote the $\alpha$ quantile of ${Y_{t,j}}$ for $\alpha \in (0,1)$. Finally, let $ \gamma_{\tau,j,j'}(\alpha):=\mathbb{P}\{(Y_{t+\tau,j}-\lambda_{j,\alpha})(Y_{t,j'}-\lambda_{j',\alpha})<0 \}$  denote the lag-$\tau$ level-crossing rate of ${Y_{t,j}}$ and ${Y_{t,j'}}$. Then, according to \cite{Li2012,Li2014}, the quantile cross-spectrum  is defined as 
\begin{equation*}
    s_{j,j'}(\omega,\alpha)=\eta_{j,\alpha}\eta_{j',\alpha}f_{j,j'}(\omega,\alpha), 
\end{equation*}

\noindent where

\begin{equation*}
    f_{j,j'}(\omega,\alpha)=\sum_{\tau=-\infty}^{\infty} \biggl \{1-\frac{1}{2\alpha(1-\alpha)}\gamma_{\tau,j,j'}(\alpha) \biggr\}\exp(i 2 \pi \omega \tau),~i=\sqrt{-1},
\end{equation*}

\noindent is called the level-crossing spectrum, and the scaling constants are defined as  $\eta_{j,\alpha}:=\sqrt{\alpha(1-\alpha)}/\dot{F}_j(\lambda_{j,\alpha})$ and $\eta_{j',\alpha}:=\sqrt{\alpha(1-\alpha)}/\dot{F}_{j'}(\lambda_{j',\alpha})$.

 Given the observations  ${Y_{t,j}}(t=1,\ldots,n)$ and quantile level $\alpha \in (0,1)$,  consider the following quantile regression problem
\begin{equation}\label{betahat}
    \widehat{\bm{\beta}}_{n,j}(\omega,\alpha):=\argmin_{\lambda_{j,\alpha} \in  \mathds{R},\bm{\beta_j} \in \mathds{R}^2} \sum_{t=1}^n \rho_{\alpha}(Y_{t,j}-\lambda_{j}-\mathbf{x}_{t}^\top (\omega) \bm{\beta}_j),
\end{equation}
where $\rho_{\alpha}(u):= u(\alpha-\mathds{I}(u<0))$ and $\mathbf{x}_{t}(\omega):=[\cos(2 \pi \omega t),\sin(2\pi \omega t)]^\top$ for $\omega \in (0,1/2)$.

Let $\widehat{\bm{\beta}}_{n,j}(\omega,{\alpha}):=[\widehat{A}_{n,j}(\omega,\alpha),\widehat{B}_{n,j}(\omega,\alpha)]^\top$ denote the quantile regression solution given by Eq. \eqref{betahat}, then, the quantile cross-periodogram between  ${Y_{t,j}}$ and ${Y_{t,j'}}$ is defined as
\begin{equation}
    q_{j,j',n}(\omega,\alpha)=n z_{j}(\omega,\alpha)z_{j'}^*(\omega,\alpha),~j,j'=1,\dots,k,
    \label{qcper}
\end{equation}
where $z_{j}(\omega,\alpha):=\frac{1}{2}\sqrt{n}{ \widehat{A}_{n,j}(\omega,\alpha)-i \widehat{B}_{n,j}(\omega,\alpha)}$. Notice that when $j=j'$ in Eq. \eqref{qcper}, it becomes the quantile periodogram of the first kind as in \cite{Li2012}. In addition, with $\alpha=0.5$ it is the  Laplace periodogram by \cite{Li2008} as a special case.  

\cite{Li2014} provided the asymptotic properties of the quantile cross-spectrum for the multivariate time series problem.  A connection between the quantile periodogram/cross-periodogram and the quantile spectra/cross-spectra was established in \cite{Li2014} (Theorem 11.3), which states that under suitable conditions, the quantile periodogram matrix $\mathbf{Q}_n(\omega,\alpha):=[q_{j,j';n}(\omega,\alpha)]$ is asymptotically distributed as $\bm{\zeta\zeta}^{\mathsf{H}}$, where $\bm{\zeta} \sim \mathcal{N}_c(\mathbf{0},\mathbf{S}(\omega,\alpha))$, with $\mathbf{S}(\omega,\alpha):=[s_{j,j'}(\omega,\alpha)]$.

 The (squared) quantile coherence  between ${Y_{t,j}}$ and ${Y_{t,j'}}$ is defined as \begin{equation}
     c_{j,j'}(\omega,\alpha)=\frac{|s_{j,j'}(\omega,\alpha)|^2}{s_{j,j}(\omega,\alpha)s_{j',j'}(\omega,\alpha)}.
     \label{qc_true}
 \end{equation} 
 It takes values between $0$ and $1$.

\subsection{VAR Spectral Model}\label{methods: YW approach} 

The VAR spectral model of order $p$ takes the form 
\begin{equation}
   \mathbf{S}_{\text{VAR}}(\omega)=\mathbf{U}_{p}^{-1}(\mathbf{\omega})~ \mathbf{V}_{p}~ \mathbf{U}_{p}^{-\mathsf{H}}(\mathbf{\omega}),
    \label{par_VAR}
\end{equation}
where
\begin{equation*}
   \mathbf{U}_{p}(\mathbf{\omega}) =\mathds{I}-\sum_{r=1}^{p}\mathbf{\Phi}(r) e^{-i2 \pi r \mathbf{\omega}}.
\end{equation*}
The model in~\eqref{par_VAR}, is the spectral matrix of the VAR process $\{\mathbf{Y}_{t}\}$ which satisfies
\citep{Priestley1981}
\begin{equation}
   \mathbf{Y}_{t}+\mathbf{\Phi}_1\mathbf{Y}_{t-1}+\ldots+\mathbf{\Phi}_p\mathbf{Y}_{t-p}=\bm{\epsilon}_t,
    \label{YW}
\end{equation}

where $ \mathbf{Y}_{t}$ is a $k$-dimensional vector and $\mathbf{\Phi}_1,\dots,\mathbf{\Phi}_p$ are $k\times k$ matrices, and $\bm{\epsilon}_t$ is a multivariate zero mean white noise process. Each of the components of the vector $\bm{\epsilon}_t$ is a univariate white noise process, uncorrelated with each other at different time points but possibly cross-correlated at common time points. 

Assuming stationarity on the VAR model, we can obtain the multivariate Yule-Walker equations by multiplying both sides of Eq. \eqref{YW} by $ \mathbf{Y}^\top_{t-h}$ and taking expectations. Denote the covariance matrix for $\mathbf{Y}_{t}$ of nonzero lag $h$ by $ \bm{\Gamma}(h)$, this gives, 

\begin{equation*}
    \bm{\Gamma}(h)+\mathbf{\Phi}_1\bm{\Gamma}(h-1)+\ldots+\mathbf{\Phi}_p\bm{\Gamma}(h-p)=\mathbf{0},
\end{equation*}
In matrix form, this can be written as
\begin{equation}
    -[\bm{\Gamma}(1) \bm{\Gamma}(2) \ldots \bm{\Gamma}(p)]=[\mathbf{\Phi}_1 \mathbf{\Phi}_2\ldots \mathbf{\Phi}_p]\widetilde{\mathbf{\Gamma}},
    \label{YW2}
\end{equation}
where
\begin{equation*}
   \widetilde{\bm{\Gamma}}=\begin{bmatrix}
\bm{\Gamma}(0) & \bm{\Gamma}(1) & \ldots & \bm{\Gamma}(p-1)\\ 
\bm{\Gamma}(1)^\top & \bm{\Gamma}(0) & \ldots & \bm{\Gamma}(p-2)\\ 
 \vdots& \vdots & \vdots & \vdots\\ 
 \bm{\Gamma}(p-1)^\top & \bm{\Gamma}(p-2)^\top & \ldots & \bm{\Gamma}(0)
\end{bmatrix} 
\end{equation*}
The Yule-Walker equations in \eqref{YW2} must be solved for $\mathbf{\Phi}_r,r=1,\ldots,p$. A brute-force approach requires the inversion of $kp\times kp$ matrices. The multivariate version of the Durbin-Levinson algorithm (see Appendix \ref{methods: Multivariate Levinson D algo}) can be used to speed up the computation. From the VAR spectral model of order $p$ in \eqref{par_VAR}, we can compute the ordinary VAR coherence by,
 \begin{equation*}
     c_{j,j'}(\omega)=\frac{|s_{j,j'}(\omega)|^2}{s_{j,j}(\omega)s_{j',j'}(\omega)}.
 \end{equation*}

\subsection{VAR estimator of quantile coherence}\label{methods: VAR approx}

The ordinary periodogram  matrix  $\mathbf{I}_n(\omega)$ for any frequency $\omega \in [0,1)$ is defined by
\begin{equation*}
    \mathbf{I}_n(\omega)=\mathbf{J}(\omega)\mathbf{J}^{*}(\omega),
\end{equation*}
where $\mathbf{J}(\omega)=n^{-1/2}\sum_{t=1}^n \left(\mathbf{Y}_t-\overline{\mathbf{Y}} \right)e^{-i2\pi \omega t  }$ and $*$ denotes the complex conjugate transpose. It is easy to show that  
    \begin{equation}
        \mathbf{I}_n(\omega)=\sum_{|h|<n} \widehat{\mathbf{\Gamma}}(h)e^{-ih2 \pi \omega},
        \label{ord_per}
    \end{equation}
where $\widehat{\mathbf{\Gamma}}(h)=n^{-1}\sum_{t=1}^{n-h}(\mathbf{Y}_{t+h}-\overline{\mathbf{Y}})(\mathbf{Y}_t-\overline{\mathbf{Y}})^\top,h \geq 0$, and $\widehat{\mathbf{\Gamma}}(h)=\widehat{\mathbf{\Gamma}}^\top(-h), h<0$. By the $2n$-point inverse discrete Fourier transform, the sample autocovariance function (ACF) $\widehat{\mathbf{\Gamma}}(h)$ can be recovered from the periodogram, as follows,
\begin{equation}
    \widehat{\mathbf{\Gamma}}(h)=(2n)^{-1}\sum_{l=0}^{2n-1} \mathbf{I}_n(\omega_l)e^{i h 2\pi \omega_l },
    \label{ord_acf}
\end{equation}

where $\omega_l=l/2n$. (See the Appendix~\ref{A3:proof} for the mathematical proof). This relationship motivates our VAR spectral estimator.

To estimate the quantile coherence, our approach involves utilizing a parametric VAR (Vector Autoregressive) spectrum of order $p$ to approximate the quantile spectral matrix $\mathbf{S}(\omega,\alpha_m)$ for a given set of equally spaced quantile levels $\alpha_m, m=1,\ldots,n_q$ within the interval $(0,1)$. We obtain the raw extended quantile periodogram matrix $\mathbf{Q}_n(\omega_l,\alpha_m)$ from the time series, where $\omega_l = l/2n$ and it is the quantile counterpart of the ordinary periodogram matrix defined in Equation~\eqref{ord_per}. The Fourier frequencies are computed in the first half and extended symmetrically. By applying the $2n$-point inverse Fourier transform to $\mathbf{Q}(\omega_l,\alpha_m)$, we obtain the Quantile Autocovariance Function (QACF), following a similar procedure as described in Equation~\eqref{ord_acf}. The QACF is computed as,

\begin{equation}
    \widehat{\bm{\Gamma}}(h,\alpha_m)=(2n)^{-1}\sum_{l=0}^{2n-1} \mathbf{Q}_n(\omega_l,\alpha_m)  e^{i 2\pi \omega_l h},~ h=0,1,\ldots,n-1.
    \label{hat_cov}
\end{equation}

We use the multivariate Durbin-Levinson algorithm to solve the multivariate Yule-Walker equations in \eqref{YW2} formed by $ \widehat{\bm{\Gamma}}(h,\alpha_m)$ in Eq. \eqref{hat_cov} for each fixed $\alpha_m$. This algorithm produces the VAR coefficients $\widehat{\mathbf{\Phi}}(1,\alpha_m),\ldots,\widehat{\mathbf{\Phi}}(p,\alpha_m)$ and the residual covariance matrix $\widehat{\mathbf{V}}_{p}(\alpha_m)$ (see Appendix \ref{methods: Multivariate Levinson D algo}). The quantile VAR spectrum of order $p$ can be expressed as

\begin{equation}
   \widehat{ \mathbf{S}}_{\text{QVAR}}(\omega,\alpha_m)=\widehat{\mathbf{U}}_p^{-1}(\mathbf{\omega},\alpha_m)~ \widehat{\mathbf{V}}_{p}(\alpha_m)~ \widehat{\mathbf{U}}_p^{-\mathsf{H}}(\mathbf{\omega},\alpha_m),
    \label{par_spec}
\end{equation}
where,
\begin{equation*}
  \widehat{ \mathbf{U}}_p(\mathbf{\omega},\alpha_m) =\mathds{I}-\sum_{r=1}^{p}\widehat{\mathbf{\Phi}}(r,\alpha_m) e^{-i2 \pi r \mathbf{\omega}}.
\end{equation*}

We propose to choose the order of the VAR model by minimizing the Akaike information criterion (AIC) \citep{Akaike1974}, i.e., 

\begin{equation*}
    \widehat{p}=\argmin_{p \in \{0,1,\ldots,p_{\text{max}}\}}\biggl\{\frac{1}{n_q}\sum_{m=1}^{n_q} n  \log|\widehat{\mathbf{V}}_{p}(\alpha_m)|+2k^2  p \biggl\},
\end{equation*}

\noindent where $n$ is the length of the time series, $k$ is the number of time series \citep{Lutkepohl2005}. Some other order selection criterion includes the Bayesian information criterion (BIC) by \cite{BIC} and the corrected AIC (AIC$_c$) by \cite{aicc}. 

Let the $j,j'$-th entry of $\widehat{\mathbf{S}}(\omega,\alpha_m)$ in \eqref{par_spec} be denoted by $\widehat{s}_{j,j'}(\omega,\alpha_m)$. Then, our preliminary parametric estimate of the quantile coherence $\widehat{c}_{j,j'}(\omega,\alpha_m)$ in  \eqref{qc_true} is defined as 

 \begin{equation}
  \widehat{c}_{j,j'}(\omega,\alpha_m)=\frac{|\widehat{s}_{j,j'}(\omega,\alpha_m)|^2}{\widehat{s}_{j,j}(\omega,\alpha_m)\widehat{s}_{j',j'}(\omega,\alpha_m)}.
  \label{par_qcoh}
\end{equation}

\subsection{Smoothing spline procedure}\label{methods: proposed smoothing}

Based on the underlying assumption that the true quantile coherence exhibits smoothness across quantiles, we propose an improvement to the parametric estimator of quantile coherence outlined in equation \eqref{par_qcoh}. Our approach involves smoothing the estimates across different quantile levels using smoothing splines for each frequency, while utilizing a shared tuning parameter.

Let the preliminary parametric quantile coherence estimate in \eqref{par_qcoh} be evaluated at $n_q$ quantile levels ${\alpha_m}(m=1,\ldots,n_q)$. At each frequency $\omega_l$, we have a one-dimensional sequence ${ \widehat{c}_{j,j'}(\omega_l,\alpha_m): m=1,\ldots,n_q}$ which we would like to smooth by smoothing splines. A simple way of doing so is to apply a standard smoothing spline procedure such as the \textsf{smooth.spline} function in R to each of these sequences independently and let the smoothing parameters be selected by the standard leave-one-out cross-validation technique. There are two potential problems with this simple method. First, using different smoothing parameters for different frequencies may introduce undesirable artifacts of discontinuity across frequencies; moreover, the standard leave-one-out cross-validation criterion is known to be ineffective for dealing with  positive correlations \citep{Altman1990,Wang1998}  which we observed in the preliminary coherence estimate across quantile levels. To overcome these difficulties, we propose a smoothing  procedure that employs a common $\lambda$ for all frequencies, so that the final  quantile coherence estimate can be expressed as

\begin{equation}
    \widetilde{c}(\omega_l,\cdot) :=\argmin_{ c(\omega_l,\cdot) } \Biggl \{ \sum_{m=1}^{n_q}[\widehat{c}(\omega_l,\alpha_m)-c(\omega_l,\alpha_m)]^2+\lambda\int_0^1 \left[\frac{\partial^2 c(\omega_l,\alpha)  }{\partial \alpha^2}\right]^2 d\alpha  \Biggl \},l=1,\ldots,n_f.
    \label{spline_proc}
\end{equation}

To select the common  smoothing parameter $\lambda$, we propose a special $\mathcal{K}$-fold cross-validation procedure. At each frequency $\omega_l$ the  sequence ${\widehat{c}_{j,j'}(\omega_l,\alpha_m):m=1,\ldots,n_q}$ is randomly split into $\mathcal{K}$ (approximately) equal-size groups. At each $\kappa=1,\ldots,\mathcal{K}$, one of the groups is reserved for testing, and the remaining $\mathcal{K}-1$ groups for training. Unlike typical $\mathcal{K}$-fold cross-validation procedure, we use the mean value of the coherence estimates in the training set denoted as $\overline{c}_{\text{pred},\kappa}(\omega_l)$,  to predict the mean value of the coherence estimates in the testing set, denoted as $\overline{c}_{\text{test},\kappa}(\omega_l)$. The smoothing parameter $\lambda$ is chosen as the minimizer of the following CV criterion 

 \begin{equation}
    \text{CV}(\lambda)=\sum_{\kappa=1}^{\mathcal{K}}  \Biggl \{ \sum_{l=1}^{n_f}[\overline{c}_{\text{pred},\kappa}(\omega_l)-\overline{c}_{\text{test},\kappa}(\omega_l)]^2\Biggl \}.
    \label{CV}
\end{equation}

The proposed CV criterion in \eqref{CV} uses the mean value prediction to deal with the correlation of the sequences across quantile levels. The standard $\mathcal{K}$-fold cross-validation criterion predicts individual values in the test set.  Because the residuals are positively correlated, the resulting tuning parameter tends to be very small for all frequencies and the level of smoothing tends to be minimal.

\section{Simulation Study}\label{sec:simualtion}
In this section, we present the results of a simulation study where the proposed method is compared with some alternatives in estimating the quantile coherence of simulated time series and the results of clustering bivariate time series based on both the quantile and the ordinary coherence.

\subsection{Simulation setup}\label{sec:simulationsetup}

The models considered in this simulation setting are
\begin{enumerate}
 \item VAR(2) model: $\mathbf{Z}_t=\mathbf{A}_1\mathbf{Z}_{t-1}+\mathbf{A}_2 \mathbf{Z}_{t-2}+\mathbf{W}_t$, $\mathbf{W}_t\sim\mathbf{\mathcal{N}}(\mathbf{0},\bm{\Sigma})$, 
 \begin{equation*}
     \mathbf{A}_1=\begin{bmatrix}
1.5 & -0.6 \\
0.3 & 0.2 \\
\end{bmatrix} ~\mathbf{A}_2=\begin{bmatrix}
-0.5 & 0.3 \\
0.7 & -0.2\\
\end{bmatrix}~\bm{\Sigma}=\begin{bmatrix}
4 & 1 \\
1 & 2\\
\end{bmatrix}
 \end{equation*}
\item VARMA(2,1) model: $\mathbf{Z}_t=\mathbf{A}_1\mathbf{Z}_{t-1}+\mathbf{A}_2 \mathbf{Z}_{t-2}+\mathbf{W}_t-\mathbf{B}_1 \mathbf{W}_{t-1}$, $\mathbf{W}_t\sim\mathbf{\mathcal{N}}(\mathbf{0},\bm{\Sigma})$
 \begin{equation*}
     \mathbf{A}_1=\begin{bmatrix}
0.816 & -0.623 \\
-1.116 & 1.074 \\
\end{bmatrix} ~ \mathbf{A}_2=\begin{bmatrix}
-0.643 & 0.592 \\
0.615 & -0.133\\
\end{bmatrix}
\end{equation*}
\begin{equation*}
\mathbf{B}_1=\begin{bmatrix}
0 & -1.248 \\
-0.801 & 0\\
\end{bmatrix}~\bm{\Sigma}=\begin{bmatrix}
4 & 2 \\
2 & 5\\
\end{bmatrix}
 \end{equation*}
\item Mixture model highly coherent at lower quantiles: the vector time series, ${\mathbf{Z}_t}=(Z_{t,1},Z_{t,2})^\top$ is a nonlinear mixture of three components given by
\begin{align*}
    \xi_1&:=\mathcal{W}_1(U_{t,1})U_{t,2}+(1-\mathcal{W}_1(U_{t,1}))U_{t,1}\\
      Z_{t,1}&:=\mathcal{W}_2(\xi_1)U_{t,3}+(1-\mathcal{W}_2(\xi_1))\xi_1,
\end{align*}
where $U_{t,1},U_{t,2}$, and $U_{t,3}$ are AR processes of mean zero and variance 1, satisfying
\begin{align*}
    U_{t,1}&=0.8 U_{t-1,1}+w_{t,1}\\
    U_{t,2}&=-0.7 U_{t-1,2}+w_{t,2}\\
    U_{t,3}&=0.55 U_{t-1,1}-0.81U_{t-2,3} +w_{t,3},
\end{align*}
where $w_{t,1}$, $w_{t,2}$, and $w_{t,3}$ are mutually independent Gaussian white noise. The mixing functions $\mathcal{W}_1(x)$ and $\mathcal{W}_2(x)$ are defined as follows
\begin{align*}
    \mathcal{W}_1(x)&=\left\{\begin{matrix}
 0.1& \text{if} & x<-0.8 \\
 0.8& \text{if}  & x>0.8 \\
\end{matrix}\right.\\
 \mathcal{W}_2(x)&=\left\{\begin{matrix}
 0.5& \text{if} & x<-0.4 \\
 0& \text{if}  & x>0.4 \\
\end{matrix}\right.,
\end{align*}
where $\mathcal{W}_1(x)$ is linear between $-0.8$ and $0.8$; $\mathcal{W}_2(x)$ is linear between $-0.4$ and $0.4$. The first component $U_{t,1}$ has a lowpass spectrum, $U_{t,2}$ has a highpass spectrum, and $U_{t,3}$ has a bandpass spectrum with a bandwidth of an AR(2) model with frequency at $0.20$. 
The second component $Z_{t,2}$ of the vector time series  ${\mathbf{Z}_t}$, is a delayed copy of $Z_{t,1}$ by 10 time units.

\item Mixture model highly coherent at higher quantiles: the vector time series ${\mathbf{Z}_t}=(Z_{t,1}, Z_{t,2})^\top$ is a nonlinear mixture of three components designed in a similar way as the previous mixture model. To obtain highly coherent higher quantiles, we just use another weighting function, $\mathcal{W}_2(x)$, which is defined as
\begin{equation*}
  \mathcal{W}_2(x)=\left\{\begin{matrix}
 0& \text{if} & x<-0.4 \\
 0.5& \text{if}  & x>0.4 \\
\end{matrix}\right.,
\end{equation*}
where $\mathcal{W}_2(x)$ is linear between $-0.4$ and $0.4$.
\end{enumerate}

In the simulation study, we consider two lengths of $n=500, 1000$ for the VAR, VARMA, and nonlinear mixtures models. For each model and length, we treat the average of 5000 raw quantile periodogram matrix as the quantile spectral matrix from which we get the  true quantile coherence.  Figure~\ref{qc_true} shows the true quantile coherence ($\mathbf{C}(\omega,\alpha):=\left[c(\omega_l,\alpha_m)\right], l=1,\ldots,n_f,m=1,\ldots,n_q$) for the VAR and VARMA models (top) and the mixture models  (bottom) with $n=500$.  We compute the quantile coherence at $93$ quantile levels,   $0.04,0.05,\dots,0.96$. We exclude extreme quantiles where the estimators may not have appropriate behaviors.

\begin{figure}[!htb]
\centering
\includegraphics[width=0.45\linewidth]{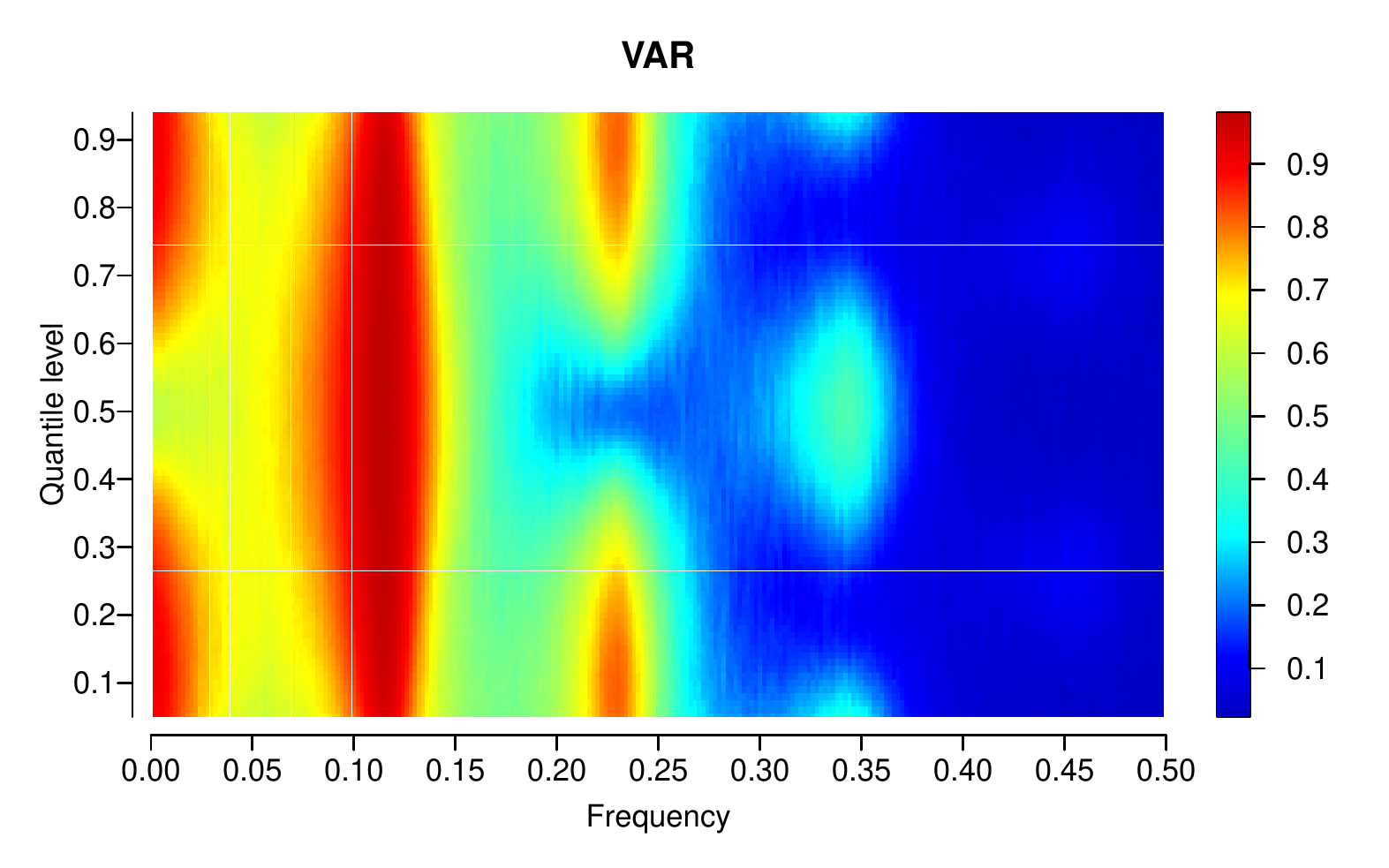}
\qquad
\includegraphics[width=0.45\linewidth]{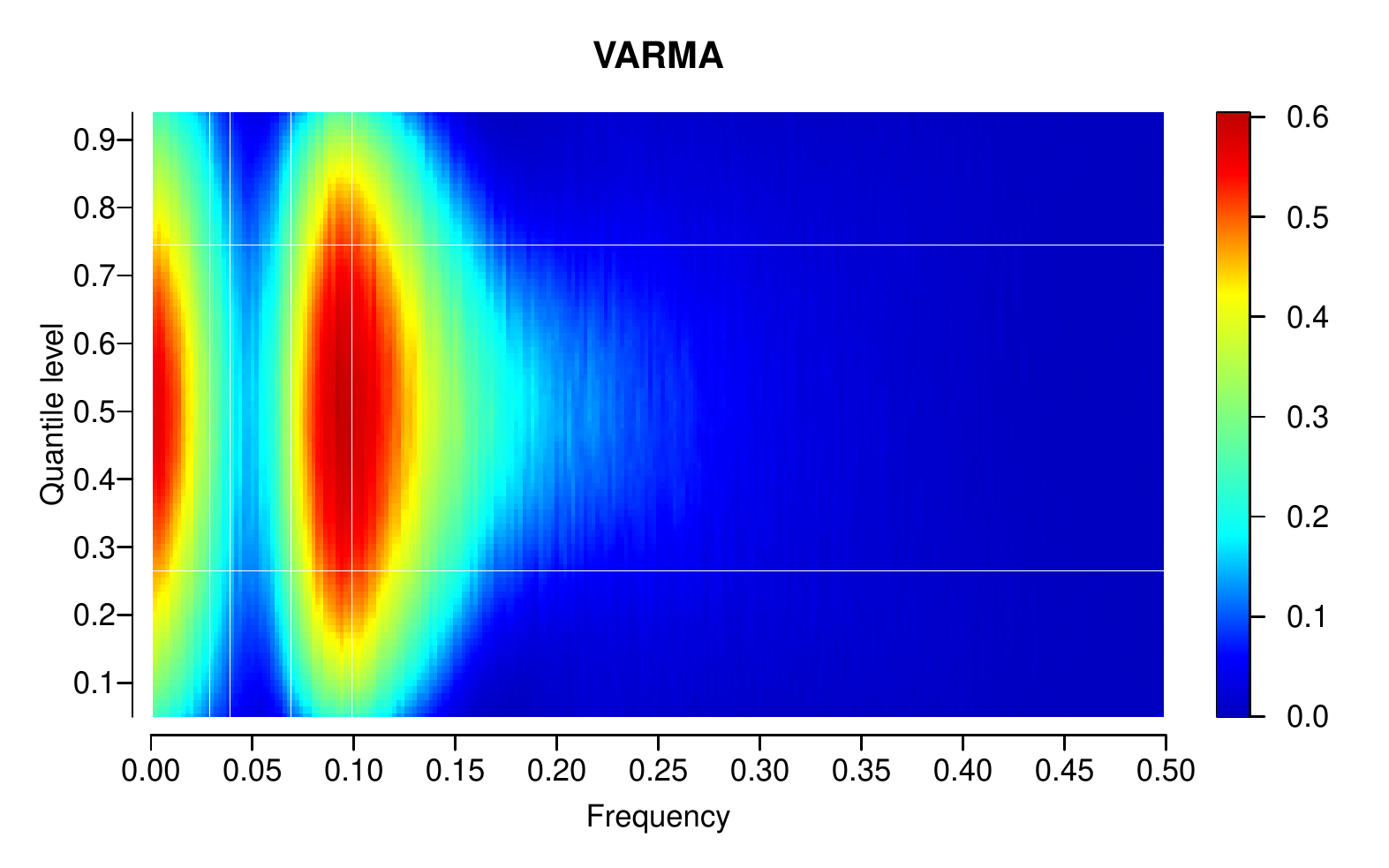}
\\
\includegraphics[width=0.45\linewidth]{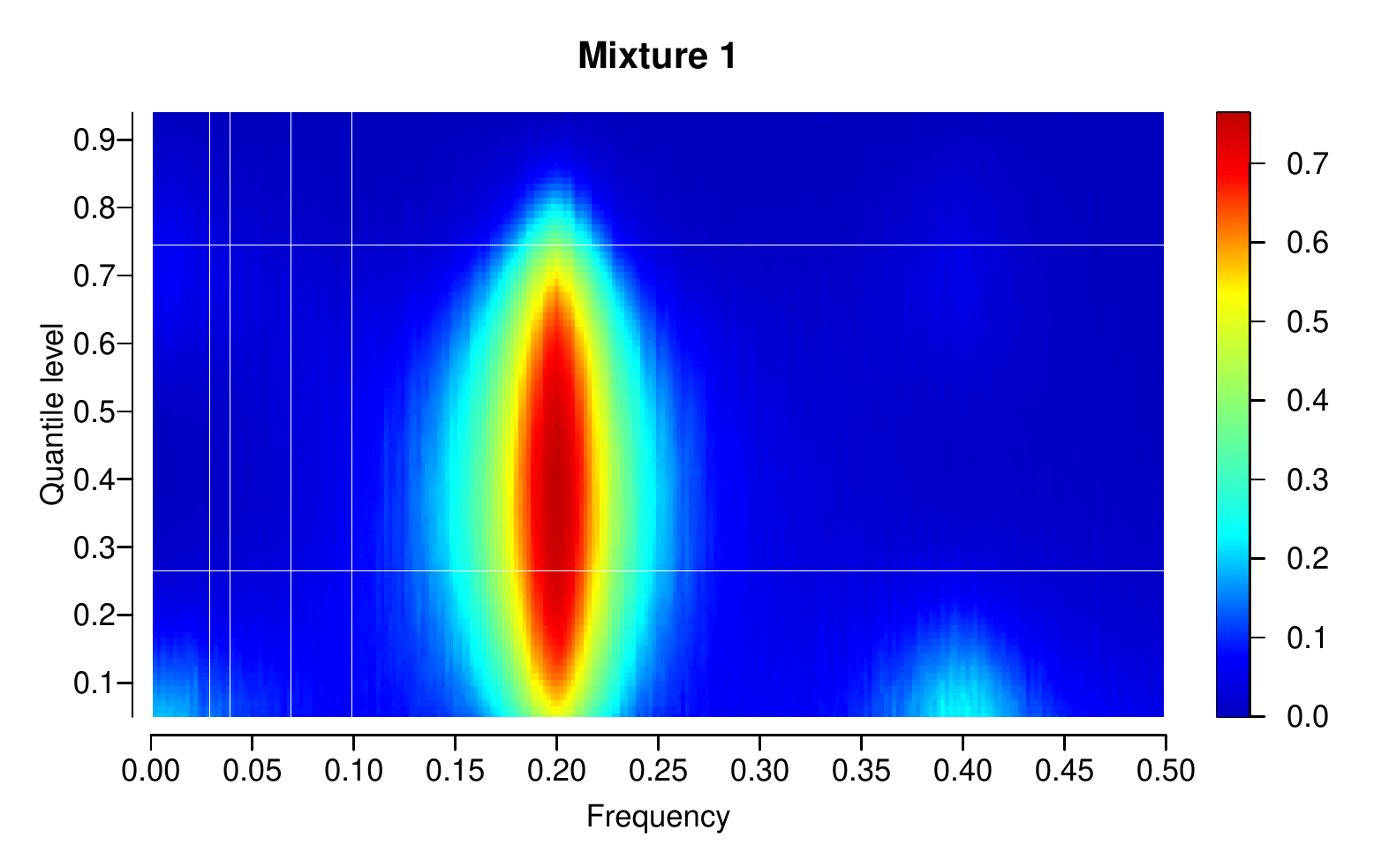} 
\qquad
\includegraphics[width=0.45\linewidth]{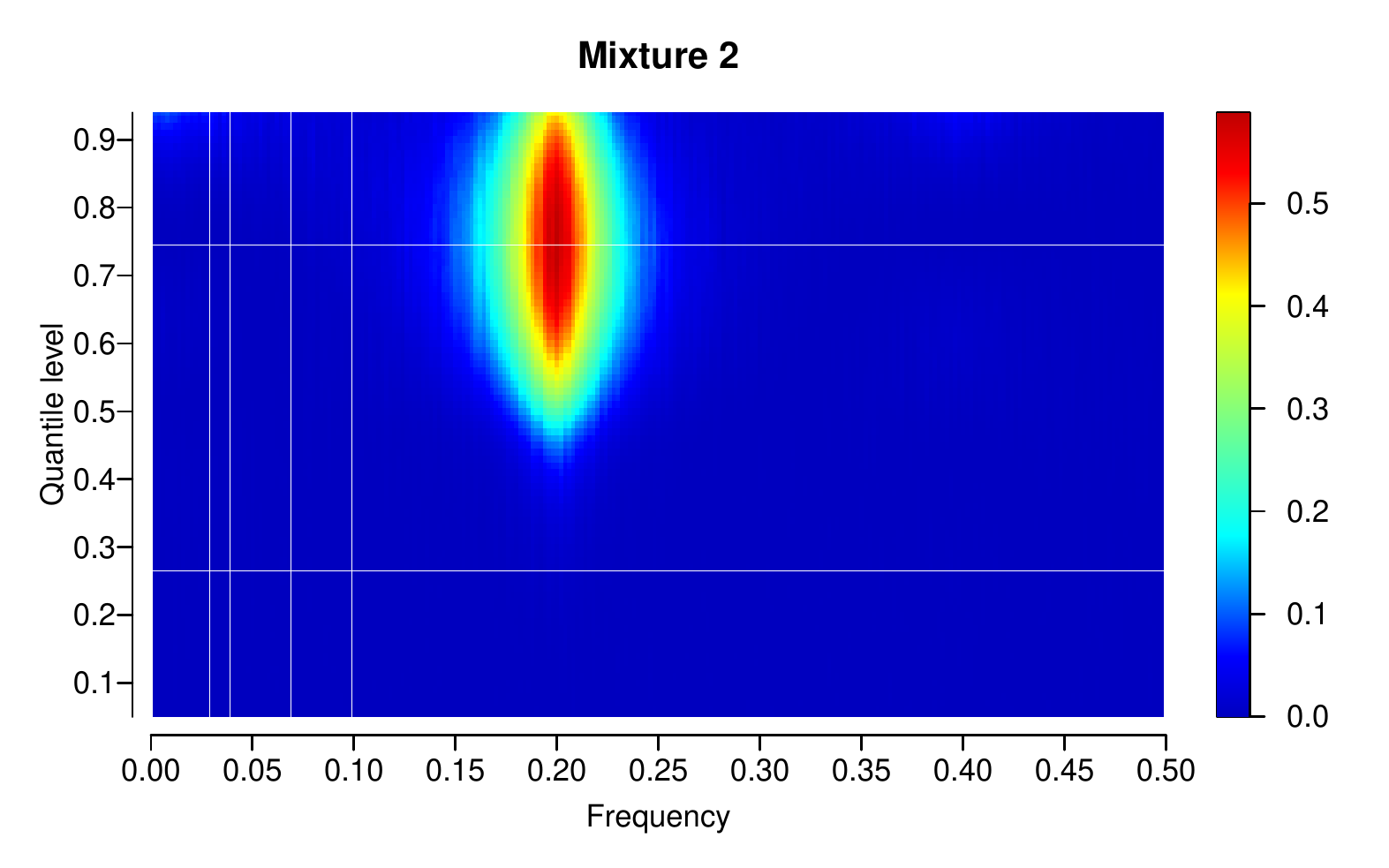}
\caption{The true quantile coherence of the 4  considered models with $n=500$ for both VAR and VARMA models (top), Mixture 1, and Mixture 2 (bottom).}
    \label{qc_true}
\end{figure} 

\subsection {Alternative Methods}

We compare our proposed estimation method for quantile coherence to three alternative methods. 

\begin{enumerate}[(a)]

    \item The nonparametric estimator proposed by \cite{QC}: we computed the quantile coherency by utilizing the matrix of smoothed rank-based copula periodograms (CCR-periodograms) across frequencies for each fixed quantile level. 
The implementation of this estimator is available in the R package \Rlogo \pkg{quantspec}  with the function \textbf{smoothedPG} \citep{Kley2016JSS}. In particular, when smoothing across frequencies, \cite{QC} uses a kernel of order $p$, where the optimal bandwidth is obtained approximately as $b_n\approx n^{-1/(2p+1))}$. For instance, in our implementations, we used the default Epanechnikov kernel, which is a kernel of order $p=2$ with an optimal bandwidth of approximately $b_n \approx n^{-1/5}$. It's important to emphasize three key distinctions between our approach and the approach presented in the study by \cite{QC}. Firstly, their estimator fixes the quantile level, whereas our estimator considers the quantile level as a continuous variable within the interval $(0,1)$. Secondly, they utilize rank-based copula periodograms, while our estimator is based on quantile periodograms obtained from trigonometric quantile regression. Lastly, their estimator relies on the selection of both kernel type and bandwidth to smooth across frequencies and does not smooth across quantiles. In contrast, our semi-parametric estimator incorporates smoothing across quantile levels by jointly selecting a common smoothing parameter.

\item The 1D smoothing spline method: we smooth the raw quantile periodogram matrix using smoothing splines, first across frequencies for each fixed quantile level, and then across quantile levels for each fixed frequency; the quantile coherence is computed based on the resulting smoothed quantile periodogram matrix.

\item The 2D kernel smoothing method: we apply 2D smoothing to the raw quantile periodograms as bivariate functions of frequency and quantile level and compute the quantile coherence from the resulting smoothed quantile periodogram matrix.

\end{enumerate}

In addition, we also demonstrate the effectiveness of the smoothing across quantiles by comparing the final estimates in \eqref{spline_proc}  which will be labeled as semi-parametric, with the preliminary estimates in \eqref{par_qcoh} which will be labeled as parametric.

We use the following root mean squared error (RMSE) between the estimated quantile coherence and the true quantile coherence to measure the performance of an estimator:
\begin{equation*}
    D_{\text{RMSE}}\{\mathbf{C},\widetilde{\mathbf{C}}\}=
    \sqrt{\frac{1}{n_f n_q}\sum_{l=1}^{n_f}   \sum_{m=1}^{n_q}[c(\omega_l,\alpha_m)-\widetilde{c}(\omega_l,\alpha_m)]^2}.
\end{equation*}

The performance assessment of different cases is presented in Table \ref{Table:Comparisons}, where the average root mean squared error (RMSE) is calculated based on 200 simulation runs. The estimation methods are labeled as follows: Semi-Param: Parametric estimator with smoothing across quantiles in \eqref{spline_proc}. The smoothing parameter is selected using a five-fold cross-validation procedure, as described in Section \ref{methods: proposed smoothing}. Parametric: VAR estimator of quantile coherence without smoothing in \eqref{par_qcoh}. BK (2019): Nonparametric estimator proposed by \cite{QC}. S.spline: 1D smoothing spline. 2D Kernel: 2D kernel smoothing.
 From the results in Table \ref{Table:Comparisons}, it is evident that the proposed semi-parametric estimator, which incorporates smoothing across quantiles, consistently outperforms all other methods. The additional smoothing significantly improves the estimation of quantile coherence overall. Moreover, even without smoothing, the parametric estimator itself demonstrates superior performance compared to its competitors. Notably, it produces better simulation results than the nonparametric estimator proposed by \cite{QC}. Furthermore, we observe a decreasing trend in both the RMSE and standard error as the sample size ($n$) increases. This suggests that the accuracy of the estimators improves with larger sample sizes.

\begin{table}[ht]
\centering
\caption{This table presents the average root mean squared error (RMSE) of the quantile coherence estimates obtained from 200 simulation runs. The minimum value in each row for each of the five estimation methods is highlighted in bold. The standard error is also provided in parentheses. }
\label{Table:Comparisons}
\resizebox{\textwidth}{!}{
\begin{tabular}{rlllllll}
  \hline
 \textbf{Model} & \textbf{n} & \textbf{Semi-Param} & \textbf{Parametric} & \textbf{BK(2019)} & \textbf{S.spline} & \textbf{2D Kernel} \\ 
  \hline
 \textbf{VAR(2)} & 500 & \textbf{0.069 (0.010)} & 0.089 (0.008) & 0.114 (0.004) & 0.451 (0.017) & 0.451 (0.017) \\ 
   \textbf{VAR(2)} & 1000 & \textbf{0.054 (0.006)} & 0.066 (0.006) & 0.104 (0.003) & 0.459 (0.014) & 0.460 (0.014) \\ 
   \textbf{VARMA(2,1)} & 500 & \textbf{0.079 (0.011)} & 0.089 (0.010) & 0.121 (0.009) & 0.389 (0.069) & 0.384 (0.058) \\ 
   \textbf{VARMA(2,1)} & 1000 & \textbf{0.059 (0.009)} & 0.065 (0.008) & 0.118 (0.006) & 0.383 (0.048) & 0.380 (0.046) \\ 
   \textbf{Mixture 1} & 500 & \textbf{0.077 (0.010)} & 0.088 (0.009) & 0.145 (0.013) & 0.445 (0.077) & 0.455 (0.062) \\ 
   \textbf{Mixture 1} & 1000 & \textbf{0.057 (0.007)} & 0.064 (0.006) & 0.146 (0.010) & 0.454 (0.067) & 0.464 (0.046) \\ 
   \textbf{Mixture 2} & 500 & \textbf{0.065 (0.010)} & 0.067 (0.011) & 0.078 (0.007) & 0.490 (0.138) & 0.531 (0.078) \\ 
   \textbf{Mixture 2} & 1000 & \textbf{0.045 (0.010)} & 0.047 (0.009) & 0.075 (0.006) & 0.466 (0.115) & 0.490 (0.072) \\ 
   \hline
\end{tabular}}
\end{table}

\subsection{Clustering simulation} \label{sec:simuationclustering}

In this section, we consider the problem of clustering bivariate $(k=2)$ time series based on the similarities of their quantile coherence. We compare this method with an alternative that employs the ordinary coherence derived from the VAR model of the bivariate time series. To derive the ordinary coherence, we first fit a VAR model to a bivariate time series and then compute the spectral matrix using the parameters from the fitted VAR model. The ordinary coherence is defined by the VAR spectral matrix in a way similar to \eqref{par_VAR}.  Through the simulation study, we would like to show the potential benefit of the quantile coherence for such a problem over the ordinary coherence. 

The clustering simulation starts by considering the four models described in Section \ref{sec:simulationsetup} as the true clusters. We simulate 200 bivariate time series from each of the four models, and then compute their quantile coherence and ordinary coherence. Both the quantile coherence and the ordinary coherence serve as dissimilarity measures for a hierarchical clustering procedure. 

For each time series, a feature vector is created by collecting the quantile coherence $c_{1,2}(\omega_l,\alpha_m)$ for $l=1,\ldots,n_f$ and $m=1,\ldots,n_q$. The dissimilarity measure for a pair of time series is defined as the Euclidean distance between the corresponding quantile-coherence-based feature vectors. A similar method is used to define the dissimilarity measure based on ordinary coherence.

Computing the dissimilarity measure for all pairs allows us to set a pairwise distance matrix, which is used as input for hierarchical clustering. The optimal number of clusters is chosen based on the so-called "elbow rule" \citep{elbow}. This rule uses the total within-cluster sum of squares (WSS) as a function of the number of clusters. 

\begin{figure}[h]
\centering
\includegraphics[width=0.45\linewidth]{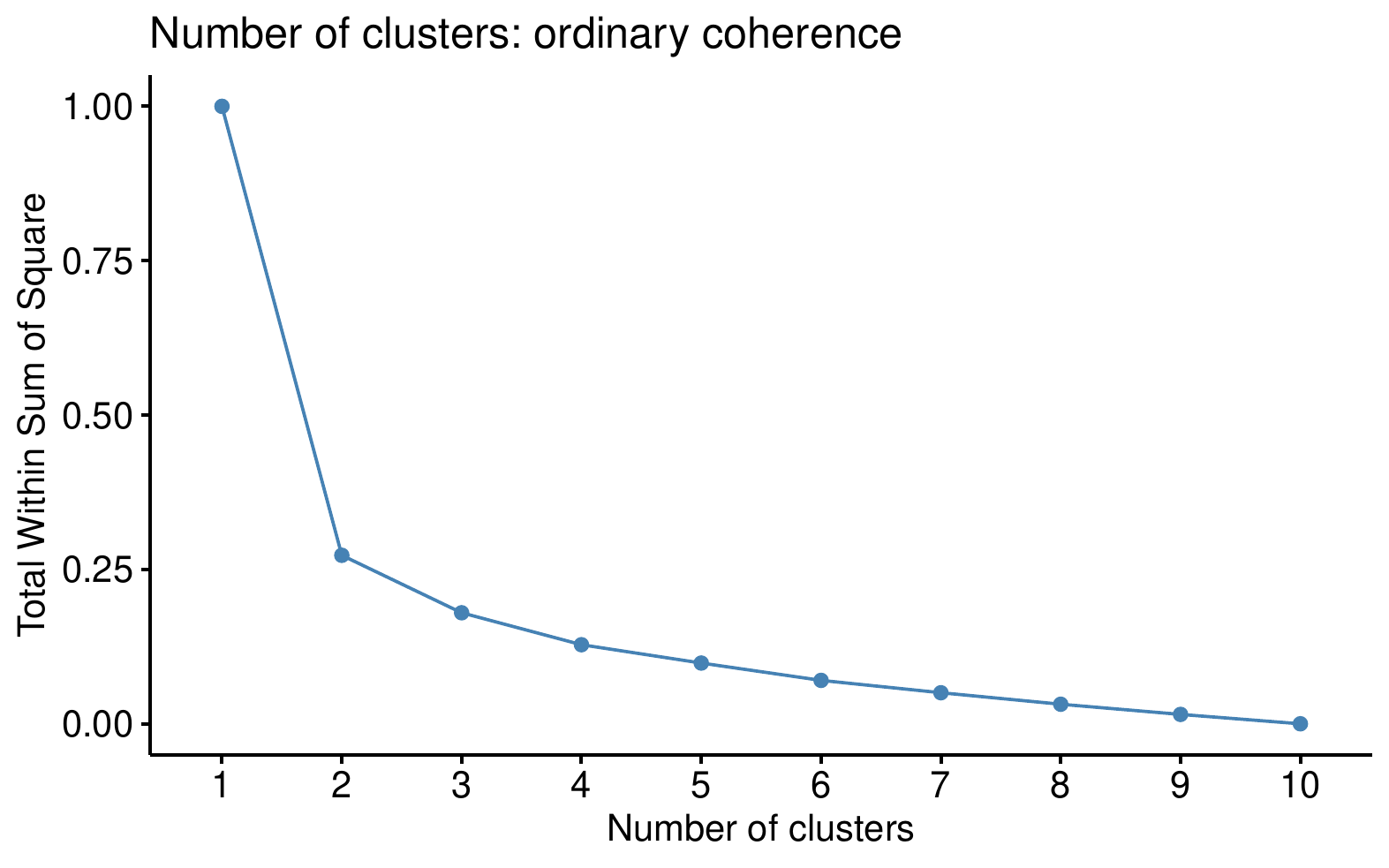}
\qquad
\includegraphics[width=0.45\linewidth]{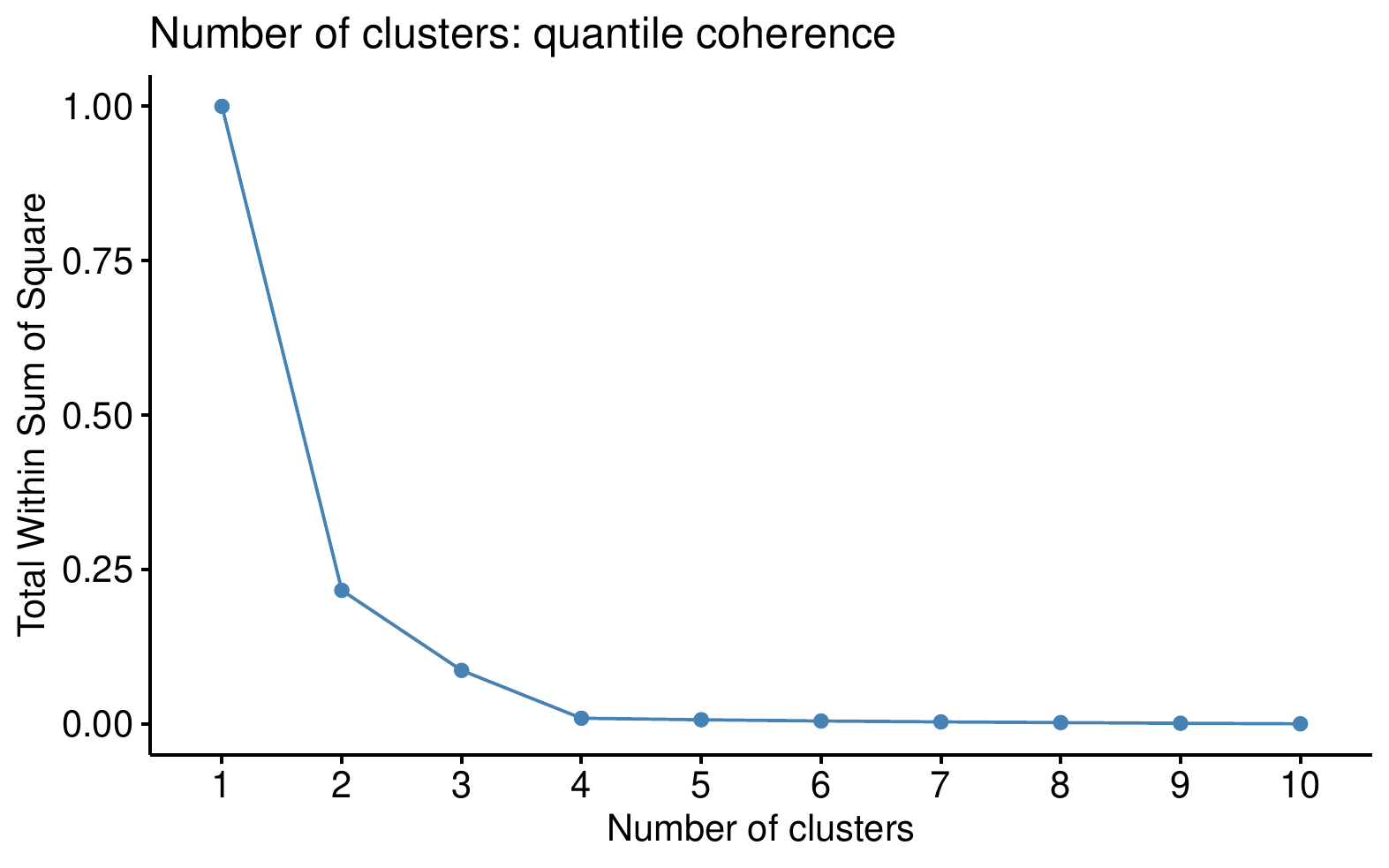}
\caption{ The total within-cluster sum of squares (WSS) as a function of the number of clusters. Based on the ordinary coherence (left), based on the quantile coherence (right). }
\label{optimal_clust_sim}
\end{figure}

From the right panel of Figure~\ref{optimal_clust_sim}, we can clearly state that the optimal number of clusters for the quantile coherence equals 4. On the other hand, the left panel of Figure~\ref{optimal_clust_sim} indicates that the optimal number of clusters for the ordinary coherence lies between 3 and 4. In a first exploratory exercise, we assumed that the number of clusters acquired by the quantile coherence, 4, was identical to the number of clusters formed from the ordinary coherence in order to evaluate the findings in terms of the allocation of members in each cluster.  

Although the optimal number of clusters is the same in this case, the allocation results differ substantially. The ordinary coherence assigns the first cluster to the 200 bivariate series acquired by model 1, likewise for the second cluster with model 2. Finally, the remaining 400 bivariate series generated by the mixture models are assigned to the third (196 members) and fourth (204 members) clusters. The third cluster contains mostly simulated bivariate time series obtained from model 3 only one case from model 4, while the fourth cluster contains members obtained mostly from model 4 and only a few cases from model 3.

In a second exercise, we chose 3 clusters as the optimal number from the ordinary coherence. The new results show that, as in the previous example, the first two clusters maintain the assignment of the bivariate series that were generated from models 1 and 2, respectively.  The difference now lies in the third cluster, which contains the 400 bivariate series obtained from the mixture models 3 and 4. By using three clusters, the ordinary coherence is not able to differentiate between the two mixture models. 
In any of the above cases, either when selecting 3 or 4 clusters, ordinary coherence is not able to correctly assign the 3 and 4 mixture models. On the contrary, the quantile coherence allows placing each of the 200 bivariate time series in a specific cluster, allowing a perfect separation of the four simulation models.

\begin{figure}[!htb]
    \centering
  \includegraphics[width=0.55\linewidth]{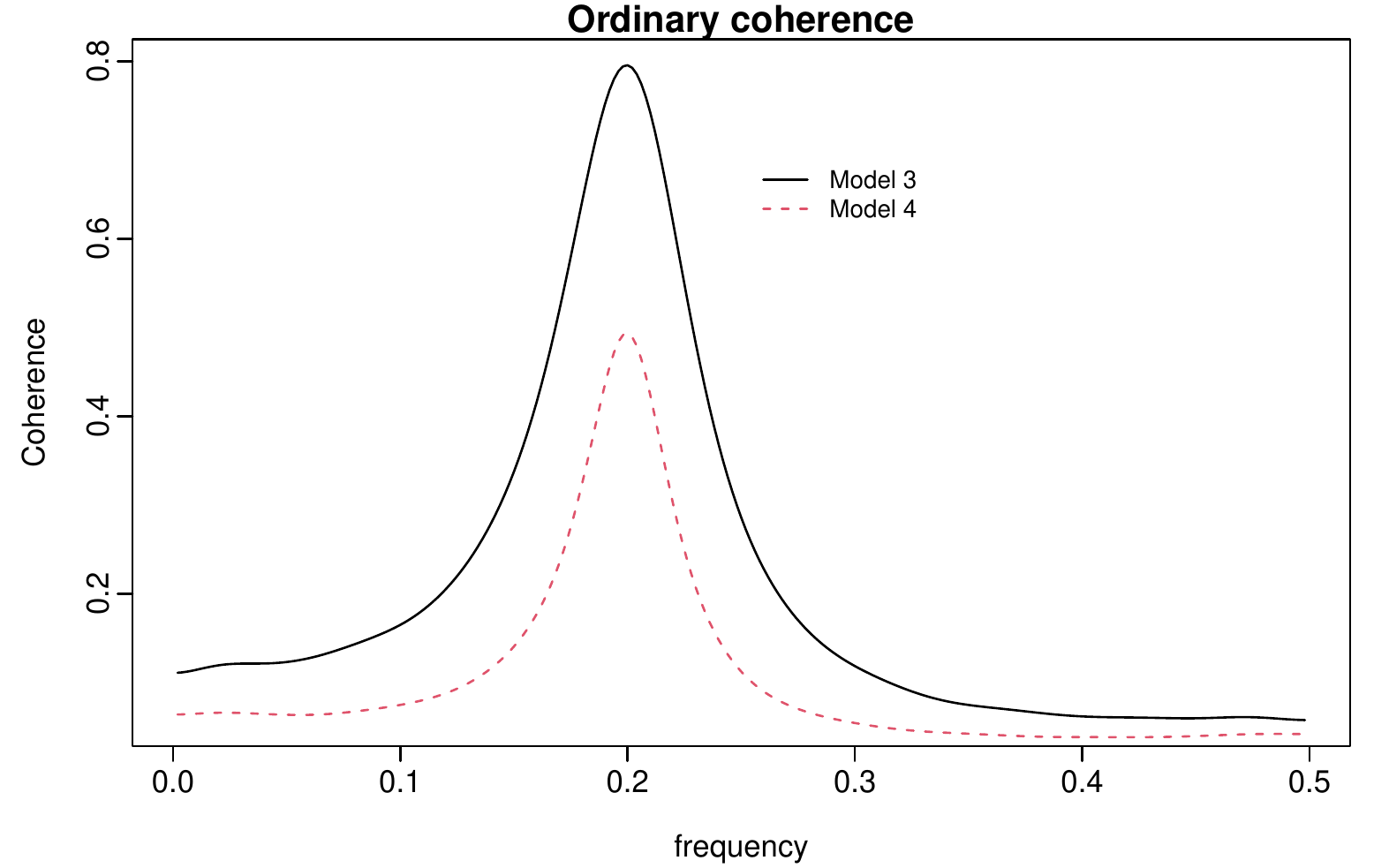}
    \caption{ The true ordinary coherence of the 2 mixture models (models 3 and 4), with $n=500$.}
    \label{fig1:exam_miss}
\end{figure}

To show some detail, we simulate the true ordinary coherence for models 3 and 4, which correspond to the mixture models. For each model, we treat the average of 5000 raw VAR periodogram matrix as the VAR spectral matrix from which we get the true ordinary coherence. In both cases, these models present a peak of coherence around frequency $0.20$ as can be seen in Figure~\ref{fig1:exam_miss}. Comparing the results of Figure~\ref{fig1:exam_miss} with those of Figure~\ref{qc_true}, we can see that the difference lies in the fact that model 3 presents a peak of coherence at frequency 0.20, but this is at low and intermediate quantile levels. On the contrary, model 4, although it presents the peak at the same frequency, is found at intermediate and high quantile levels.  Compared to the ordinary coherence, the quantile coherence offers more accurate clustering since it contains additional information about quantile levels that can be used to assign bivariate series, especially with mixture models appropriately. 

\section{Financial time series clustering}\label{sec:application}

In this section, our goal is to explore the advantages of clustering the time series of stock prices based on their quantile coherence with a benchmark. The objective of this experiment is to determine if meaningful clusters can be identified among these stocks by considering their co-variability with respect to the benchmark across different quantile regions. We selected 52 stocks from the S\&P 500 (SPX) to evaluate the estimation method for quantile coherence and its effectiveness for clustering purposes.

The 52 selected stocks represent different market sectors such as Health Care, Technology, Materials, Consumer Staples, Industrials, Consumer Discretionary, Commodities, Entertainment, Energy, Agricultural Products, Communications Services, Utilities, and Restaurants. Furthermore, the study period is 2010 to 2019, corresponding to a period free of large oscillations like the "Great Recession" and the COVID-19 pandemic.

In this experiment, the time series data consists of the daily log returns of the closing prices of specific companies within the SPX index. Each individual series is represented by a feature vector that incorporates the quantile coherence between the series and the SPX index. This quantile coherence is evaluated at Fourier frequencies ranging from 0 to 1/2. Additionally, the feature vector includes quantile regions, which are subsets of the range from 0.04 to 0.90, totaling 90 quantile levels. To quantify the dissimilarity between these series, a dissimilarity matrix is constructed. This matrix is based on the pairwise Euclidean distances calculated from the feature vectors derived from quantile coherence.

The proposed approach provides a notable advantage in accurately capturing the combined behavior of log returns with the SPX index. This emphasizes the significance of integrating information about the quantile-dependence characteristics of the data, which holds valuable insights for risk analysis and portfolio construction. In practice, portfolio managers commonly prioritize a fixed quantile level, usually below $10\%$, as it offers a simplified implementation compared to using a range of quantiles. Specifically, our analysis focuses on three quantile regions: the lower quantile region (quantiles from $4\%$ to $10\%$), the middle quantile region (quantiles ranging from $40\%$ to $55\%$), and the upper quantile region (quantiles from $70\%$ to $96\%$). The results of the cluster assignments obtained from the hierarchical clustering procedure are presented in Table~\ref{Tab:quantile_region}, while the corresponding dendrograms can be found in Appendix~\ref{A2:dend}.

\begin{table}[]
\resizebox{\textwidth}{!}{
\begin{tabular}{lll}
\toprule
\textbf{Quantiles} & \textbf{Clusters}  & \textbf{Stock Members}  \\
\midrule
\multirow{3}{*}{\textbf{Lower}} & \multirow{2}{*}{\textbf{Cluster 1}} &  ADI AFL AIG ALL AON APD C CSCO DIS DOV EMN    \\
& & IBM JPM LOW MMC MMM MSFT SBUX WFC\\ \cmidrule{2-3}
                                          & \multirow{2}{*}{\textbf{Cluster 2}} & AAPL ABT ADM AES AEE AEP AMZN CMCSA CSCO CVX F  \\ 
                                          & &  KMB KO M MCD MMM MSFT PEP PFP T TM WEN WMT XOM \\ \cmidrule{2-3}
                                          & \textbf{Cluster 3} & CPB ENB GIS MO NEM \\
                                          \midrule
\multirow{3}{*}{\textbf{Middle}}& \multirow{2}{*}{\textbf{Cluster 1}} & AFL ADBE ADI AIG ALL APD AON C CSCO CVX DIS \\
& & DOV EMN IBM JPM MMM MMC MSFT SBUX WFC XOM \\
\cmidrule{2-3}
                                          & \multirow{2}{*}{\textbf{Cluster 2}} &  AAPL ABT ADM AES AMGN AMZN BSX CL CMCSA     \\
                                          & &  F GD KMB KO LOW M MCD PEP PFP PG T TGT TM WMT \\ \cmidrule{2-3}
                                          
                                          & \textbf{Cluster 3} & AEE AEP CPB ENB GIS MO NEM WEN     \\
                                          \midrule
\multirow{3}{*}{\textbf{Upper}} & \multirow{2}{*}{\textbf{Cluster 1}} & ADBE AFL AIG APD C CSCO CMCSA DIS DOV EMN   \\
&& IBM JPM MMC MMM MSFT WFC \\ \cmidrule{2-3}
                                          
                                          & \multirow{2}{*}{\textbf{Cluster 2}} & AAPL ABT ADM AES AON AMGN AMZN BSX CL CVX F GD \\ 
                                          &&  KO LOW M MCD PEP PFP PG SBUX T TGT TM WMT XOM \\ \cmidrule{2-3}
                                          & \textbf{Cluster 3} & AEE AEP CPB ENB GIS KMB MO NEM WEN
\\
                                          \bottomrule
\end{tabular}}
\caption{Stock assignment in hierarchical clustering based on specific quantile regions}
\label{Tab:quantile_region}
\end{table}

\begin{figure}[h]
\centering
\includegraphics[width=0.45\textwidth]{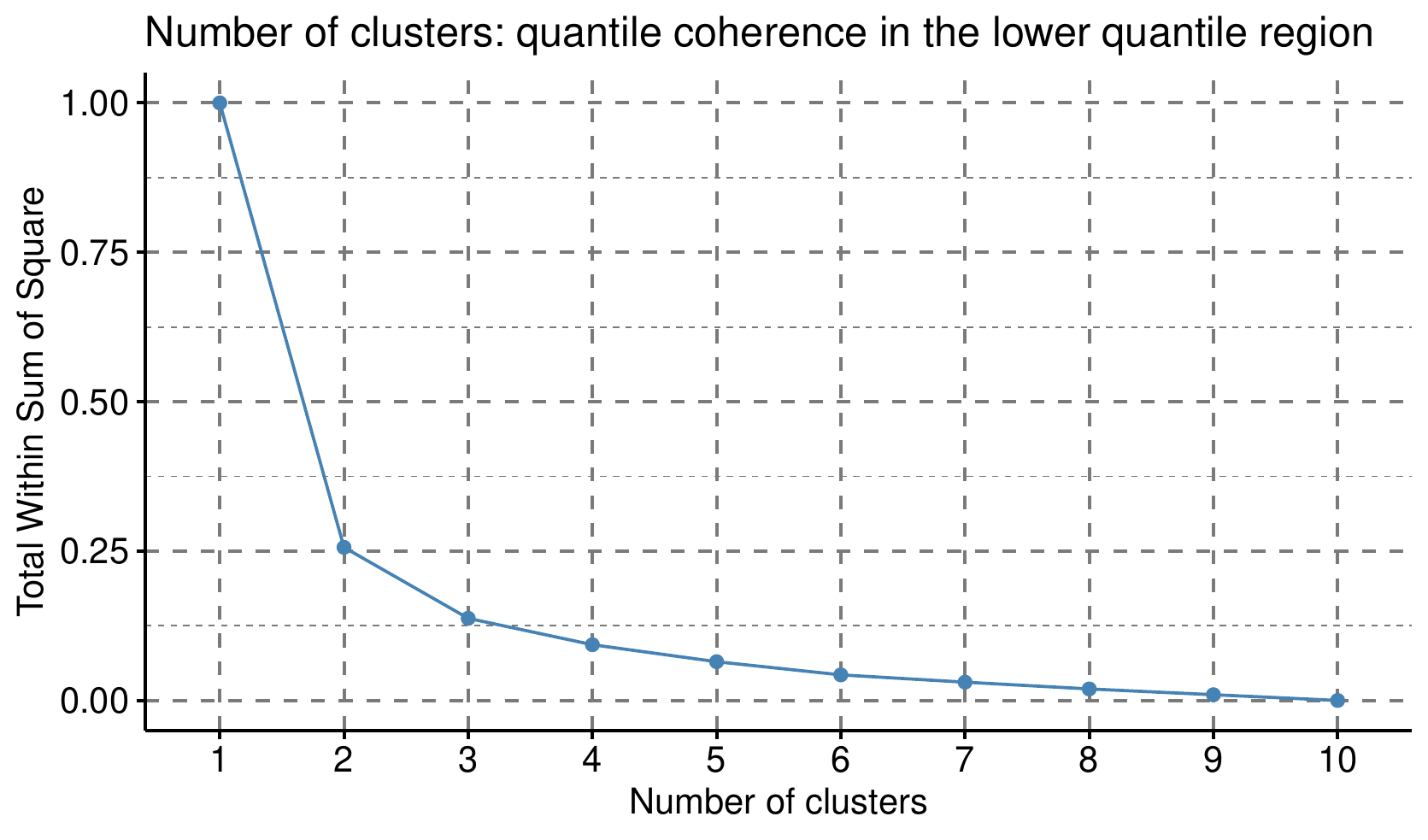}
\\
\includegraphics[width=0.45\textwidth]{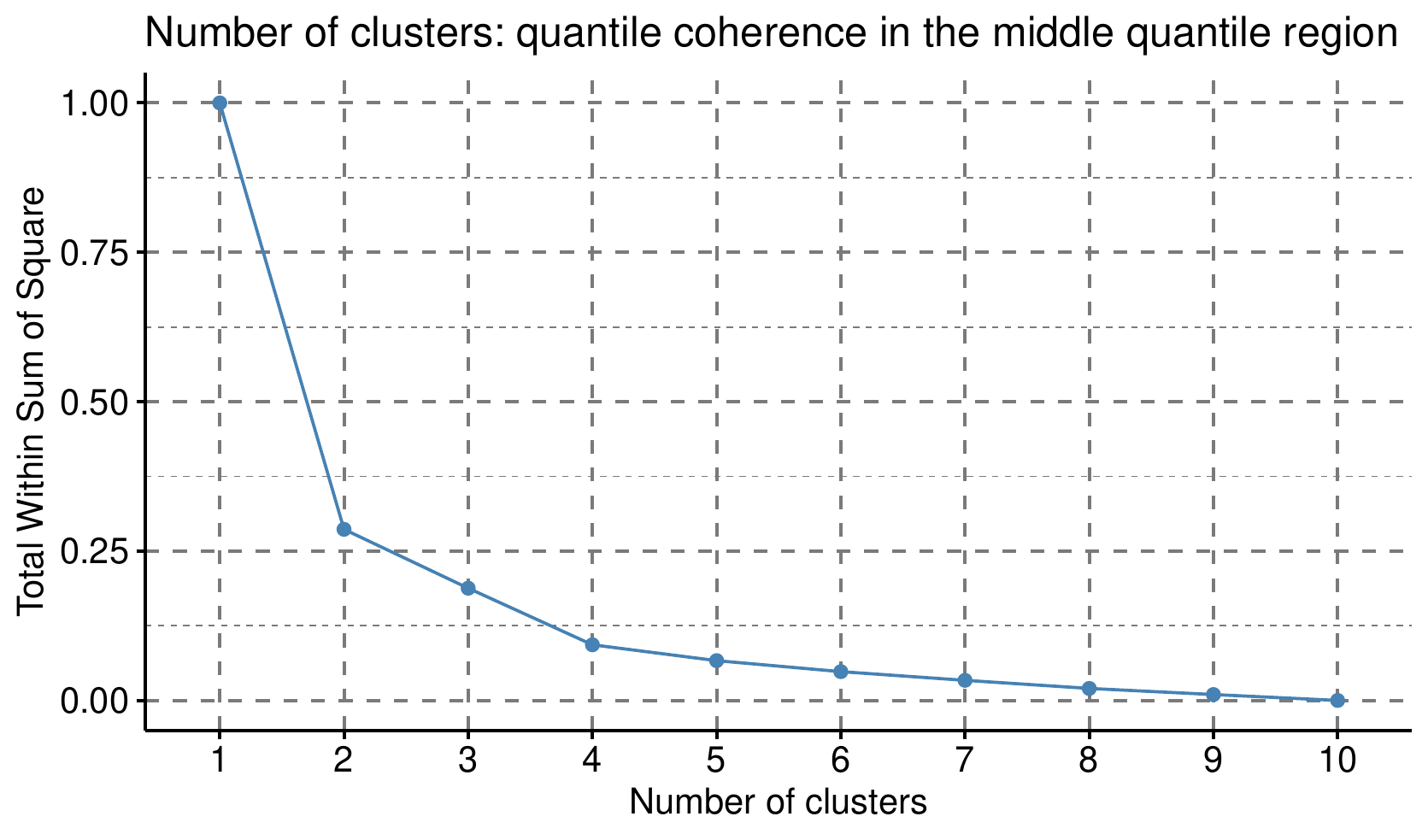}
\\
\includegraphics[width=0.45\textwidth]{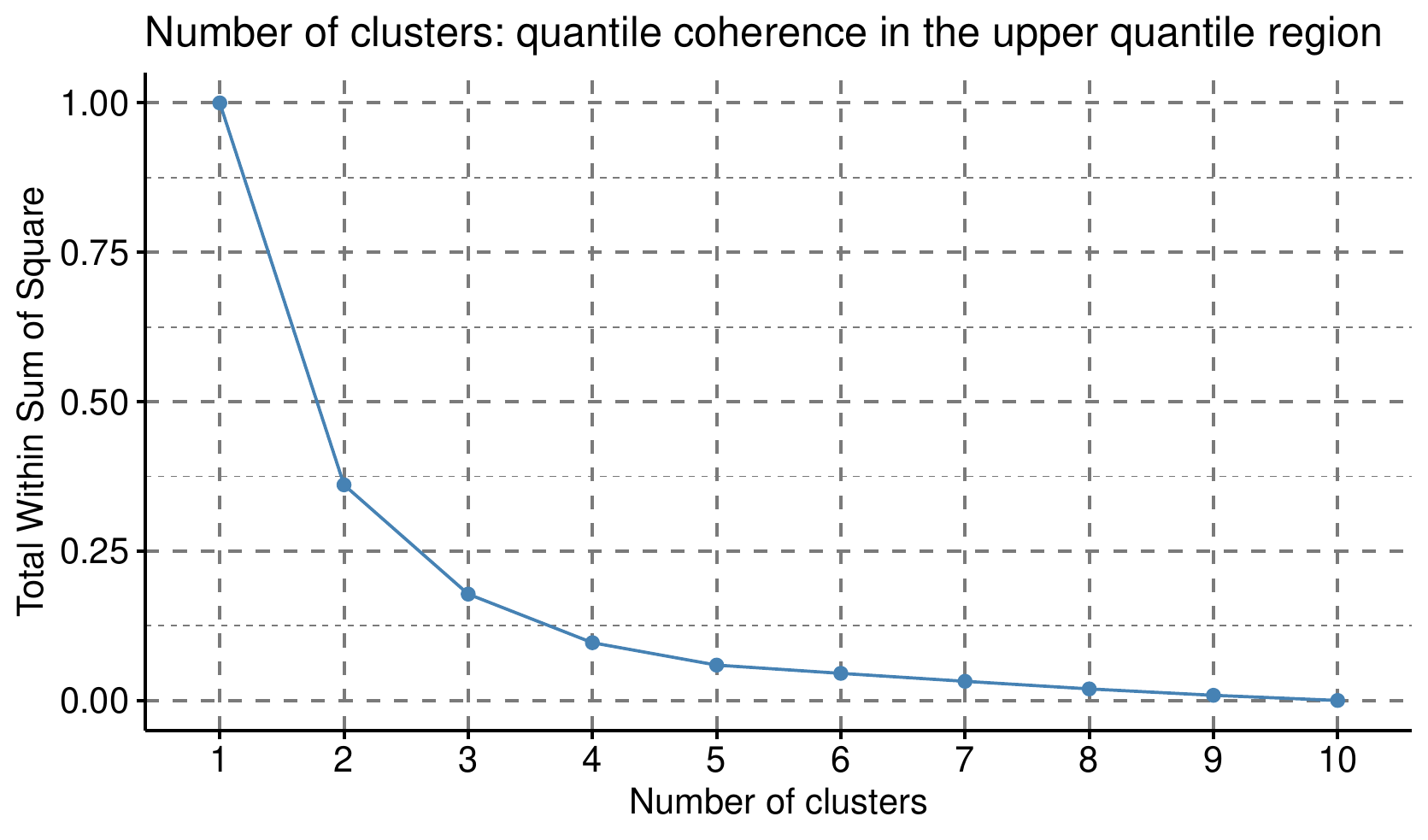} 
\caption{ The total within-cluster sum of squares (WSS) as a function of the number of clusters. (Top) clusters formed using quantile coherence for lower quantile levels. (center) clusters based on quantile coherence for middle quantile levels. (Bottom)  clusters based on quantile coherence for upper quantile levels.  }
\label{Application:elbow_upper_lower}
\end{figure}

As mentioned in Section~\ref{sec:simuationclustering}, we determine the optimal number of clusters from the dendrogram using the "elbow rule" \citep{elbow}. The within-cluster sum of squares (WSS) is depicted in Figure~\ref{Application:elbow_upper_lower}, with the top panel representing lower quantiles, the center panel representing middle quantiles, and the bottom panel representing upper quantiles. In all cases, the optimal number of clusters is determined to be 3.

\begin{figure}[!ht]
\centering
\includegraphics[width=4.8cm]{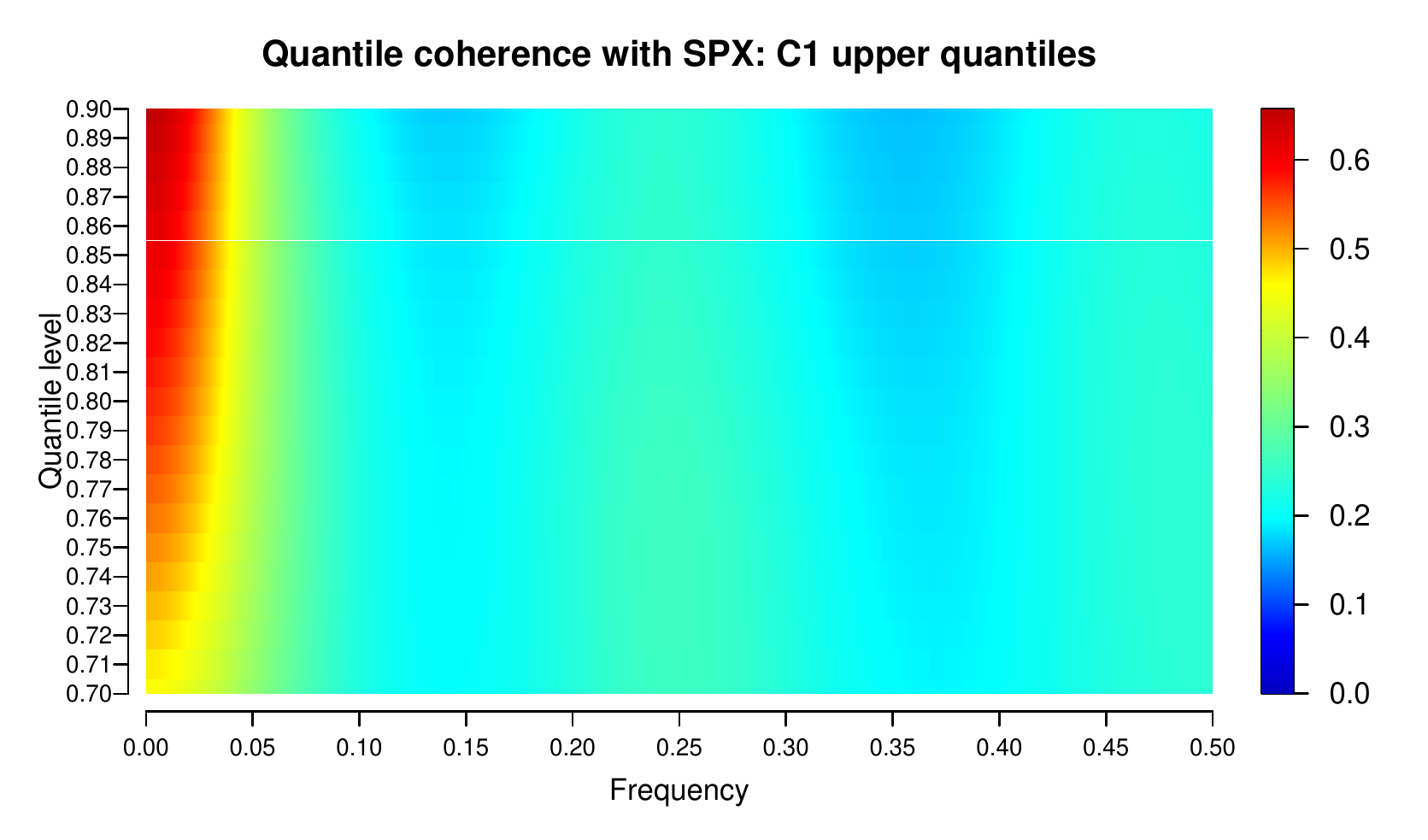}
\qquad
\includegraphics[width=4.8cm]{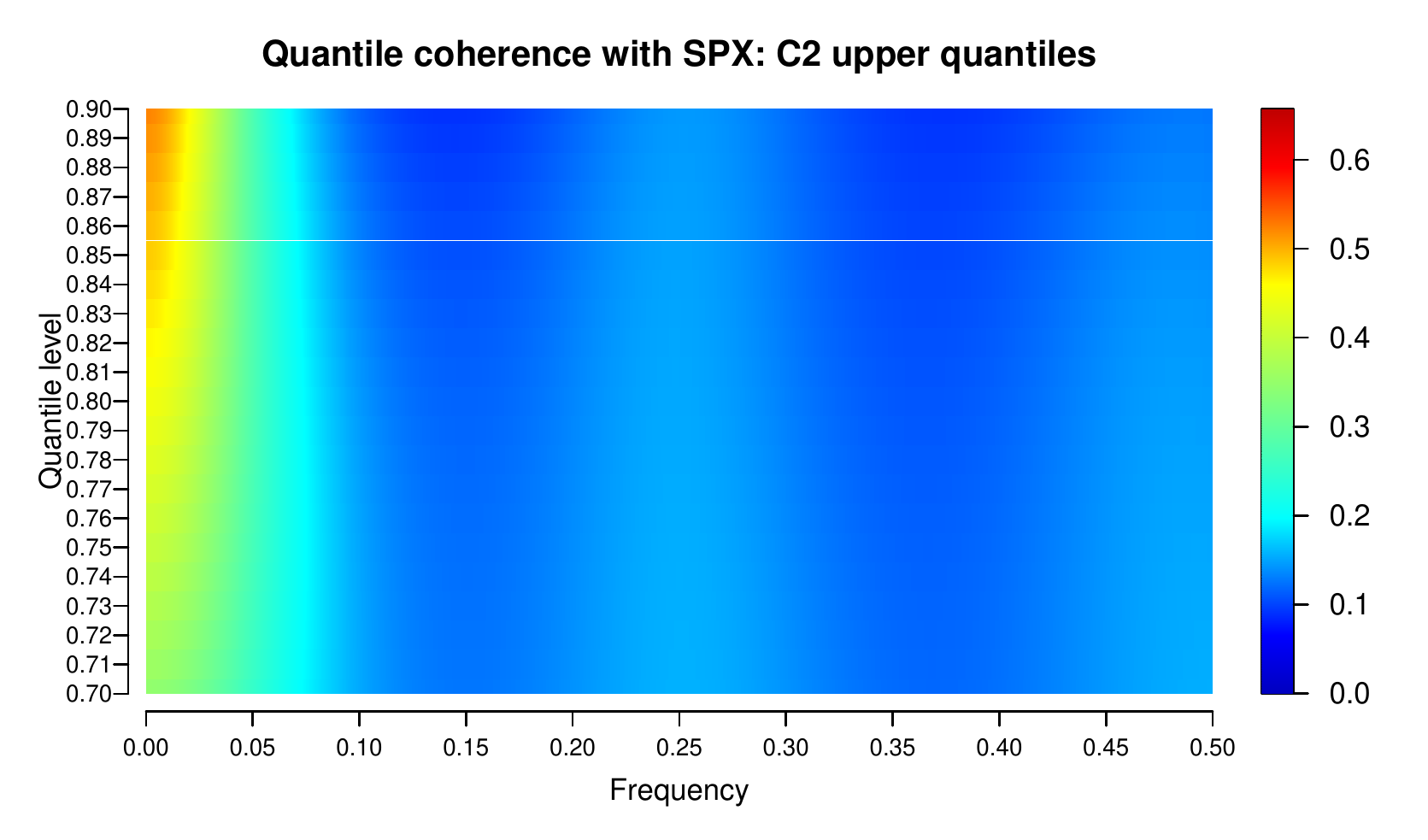}
\qquad
\includegraphics[width=4.8cm]{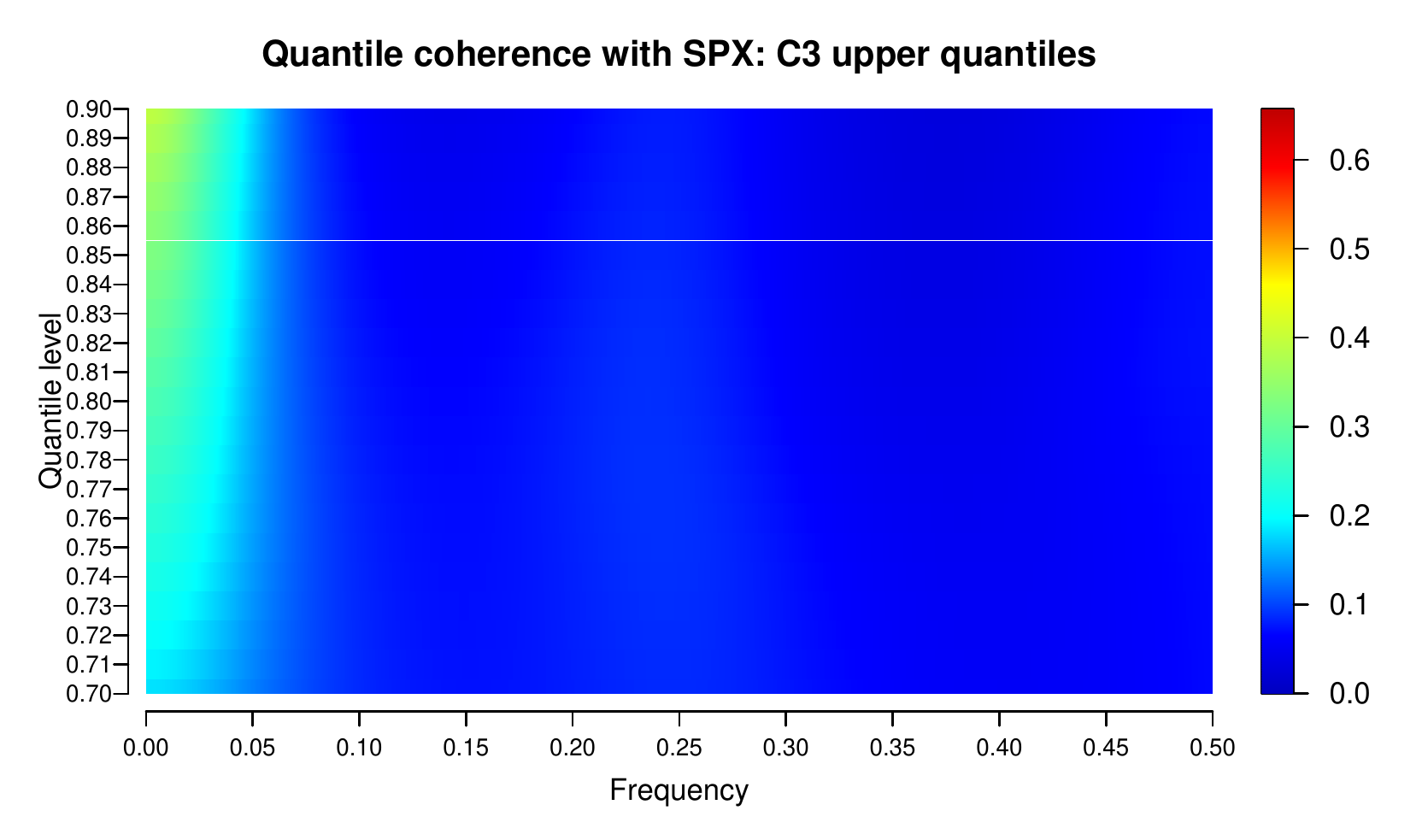}
\\
\includegraphics[width=4.8cm]{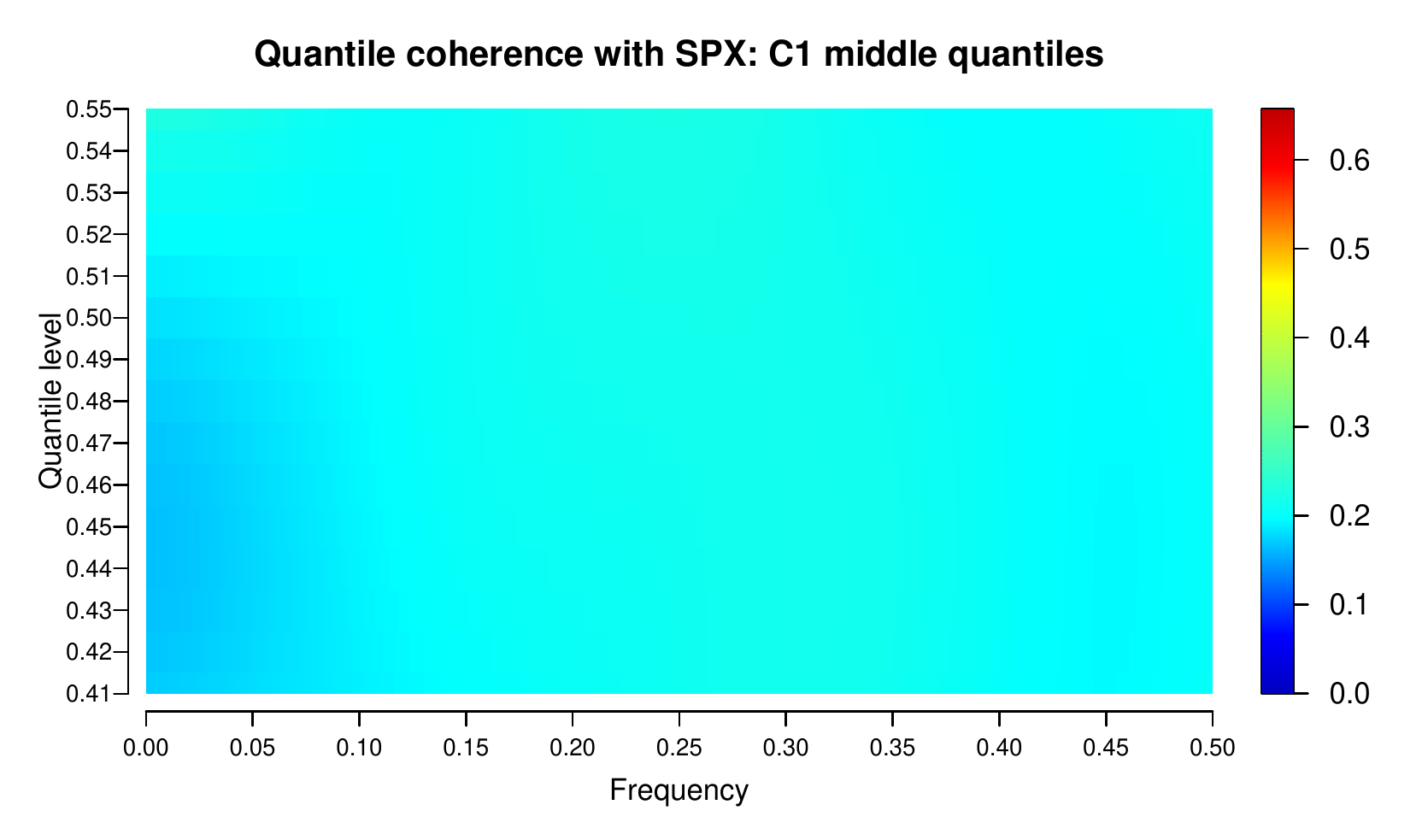}
\qquad
\includegraphics[width=4.8cm]{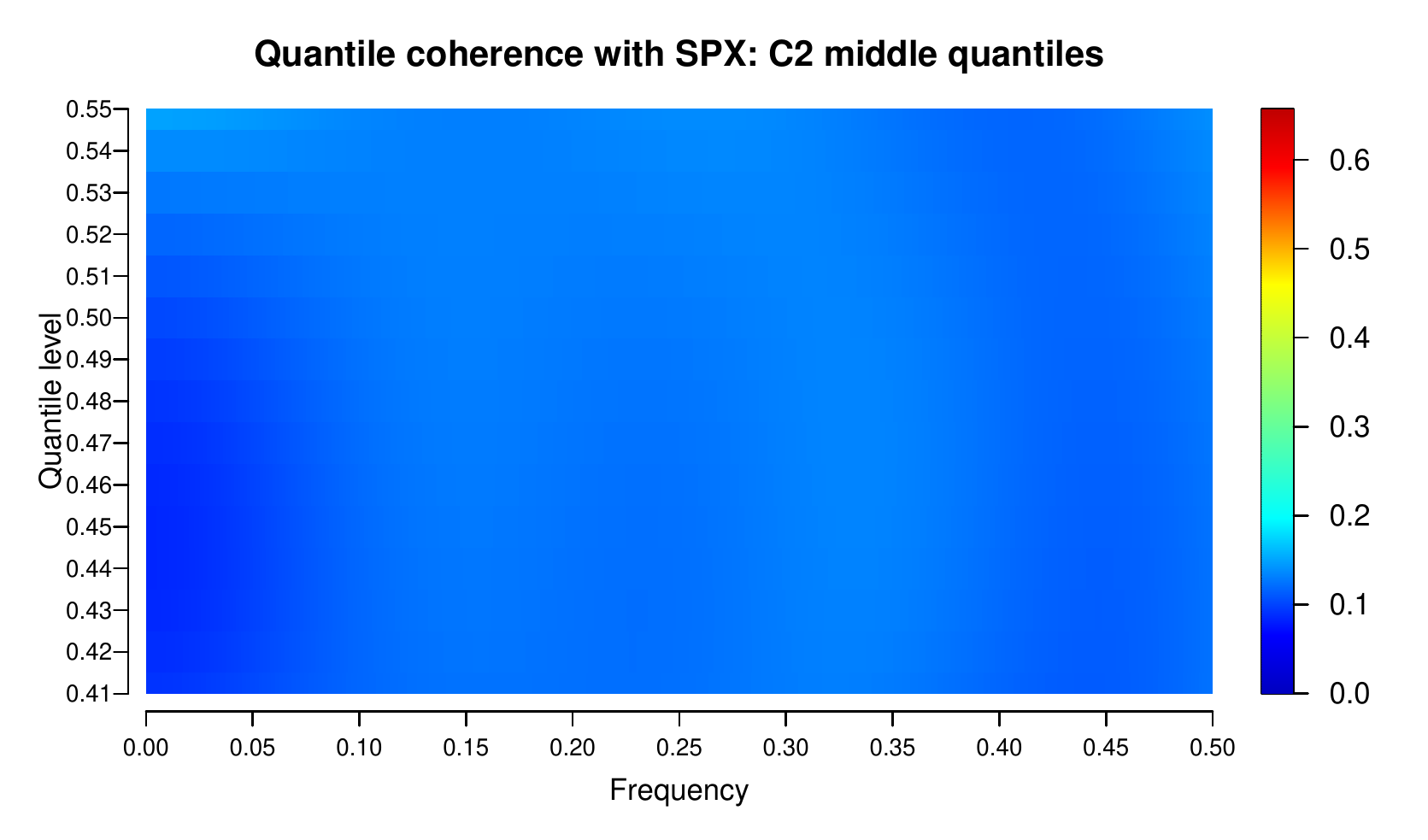}
\qquad
\includegraphics[width=4.8cm]{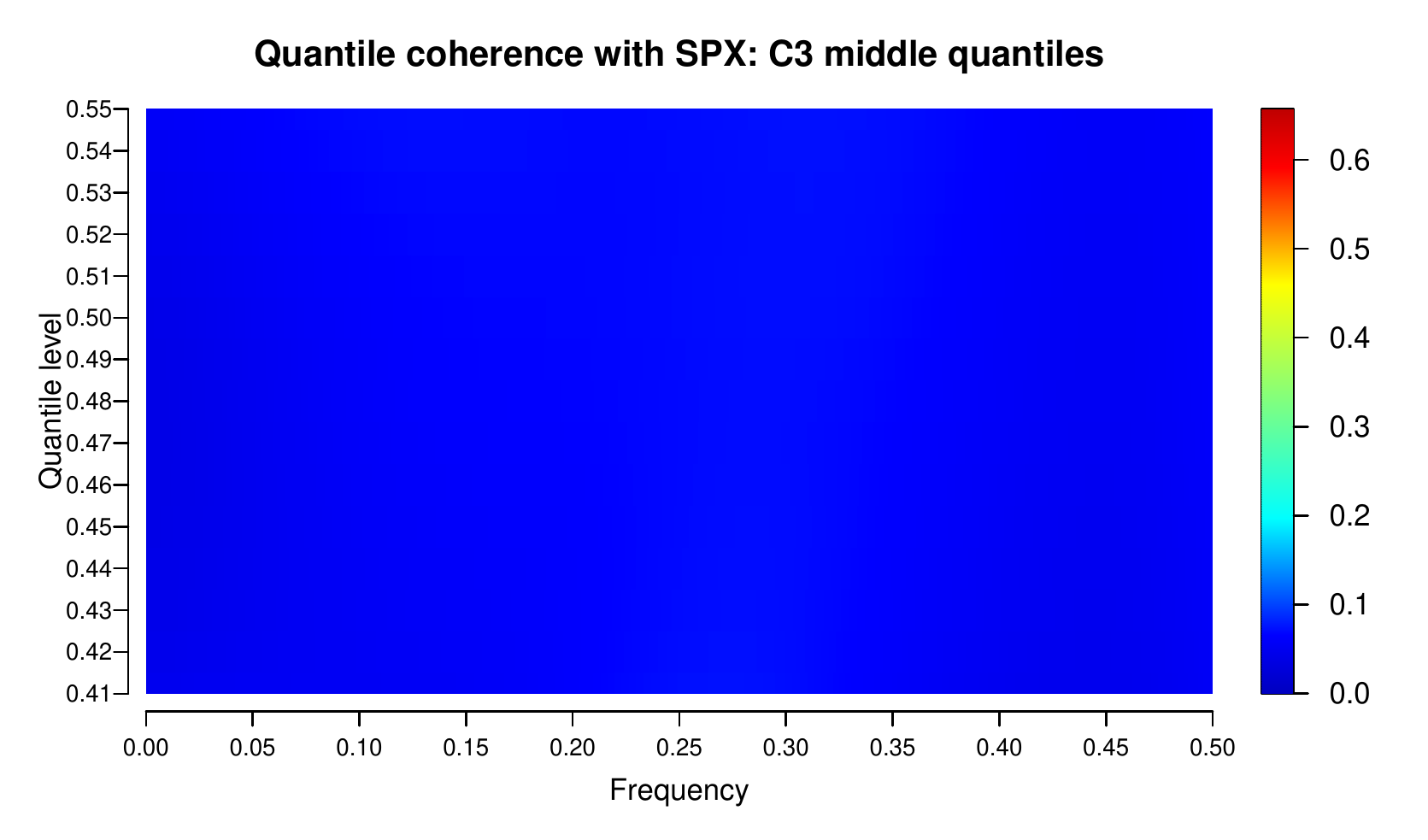}
\\
\includegraphics[width=4.8cm]{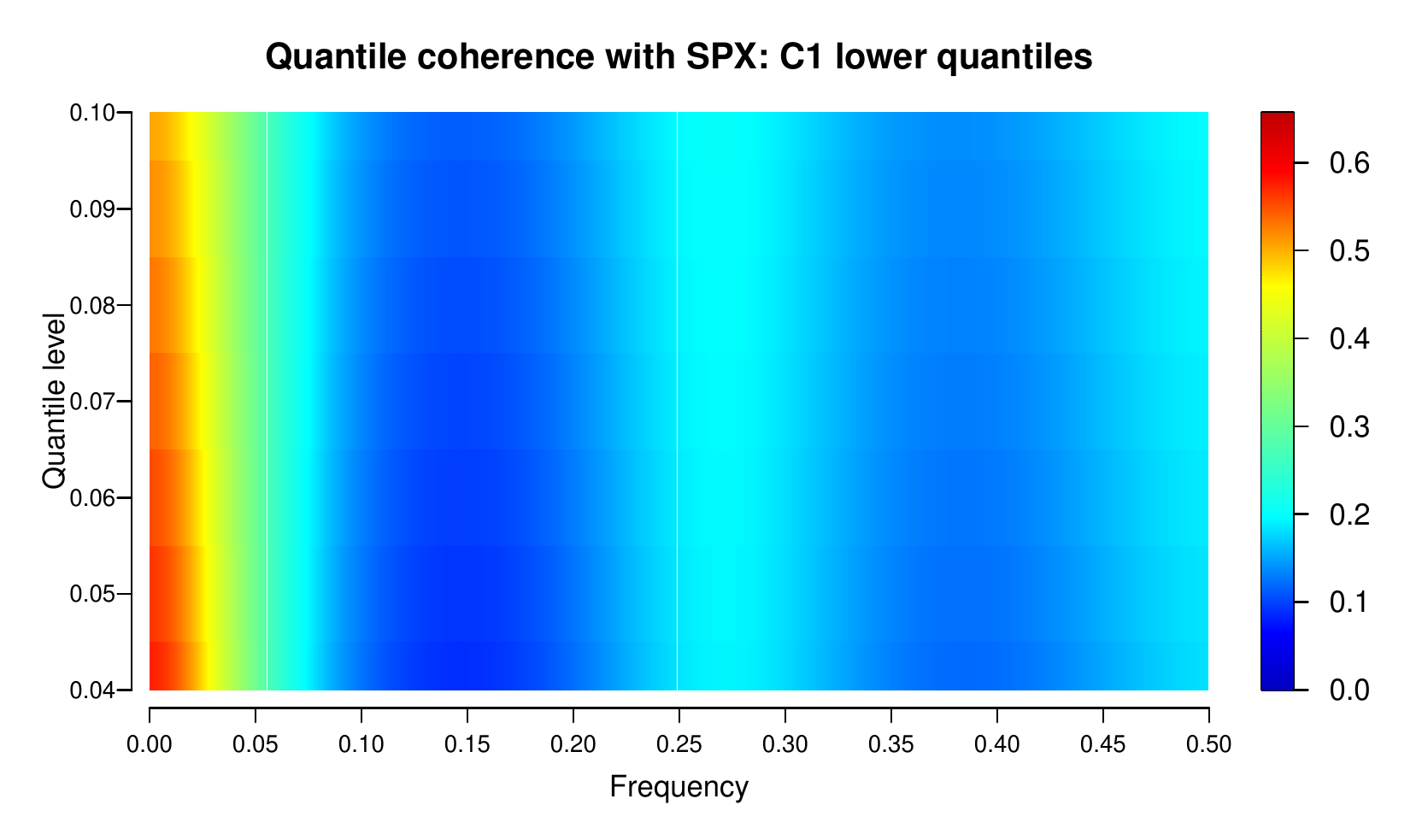}
\qquad
\includegraphics[width=4.8cm]{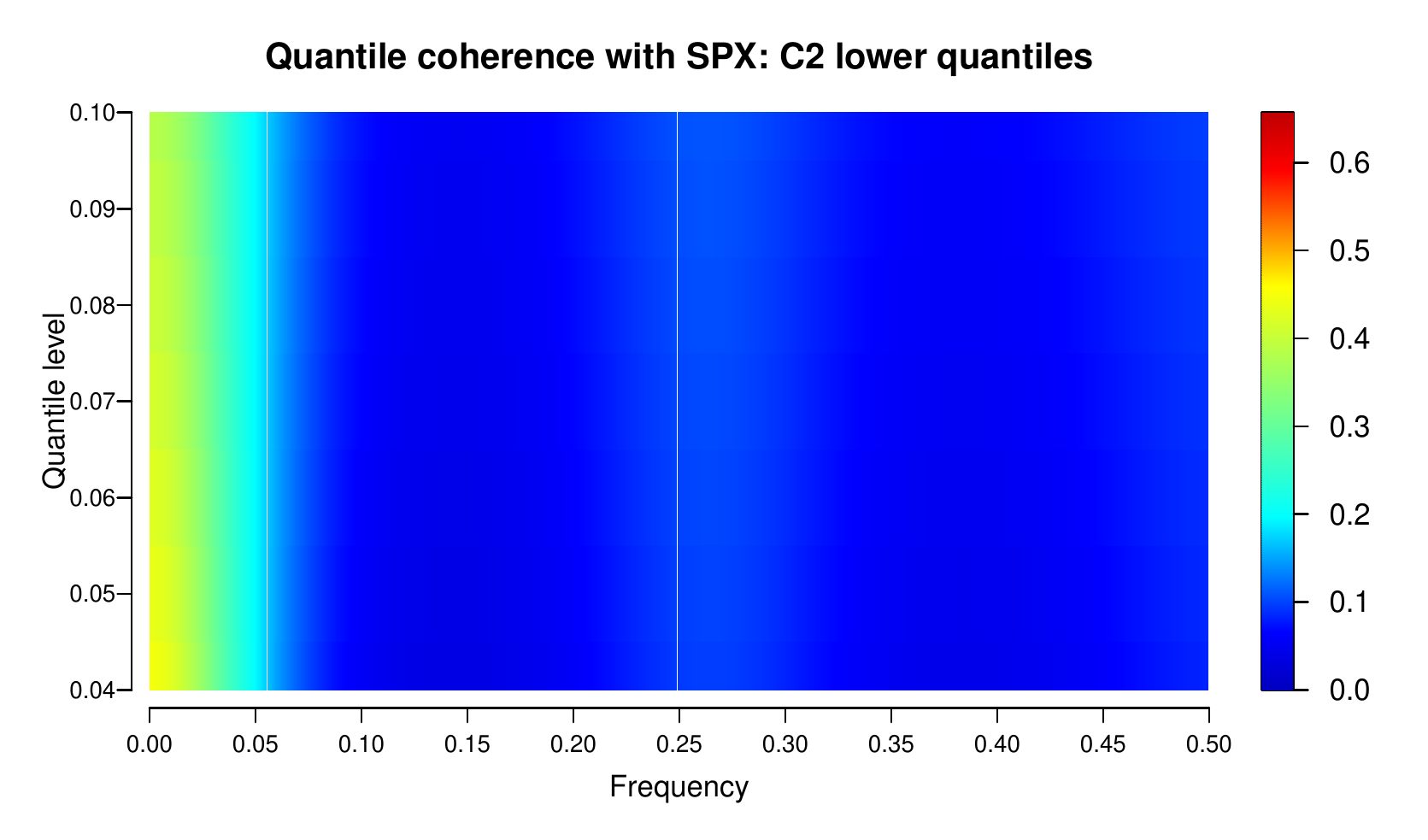}
\qquad
\includegraphics[width=4.5cm]{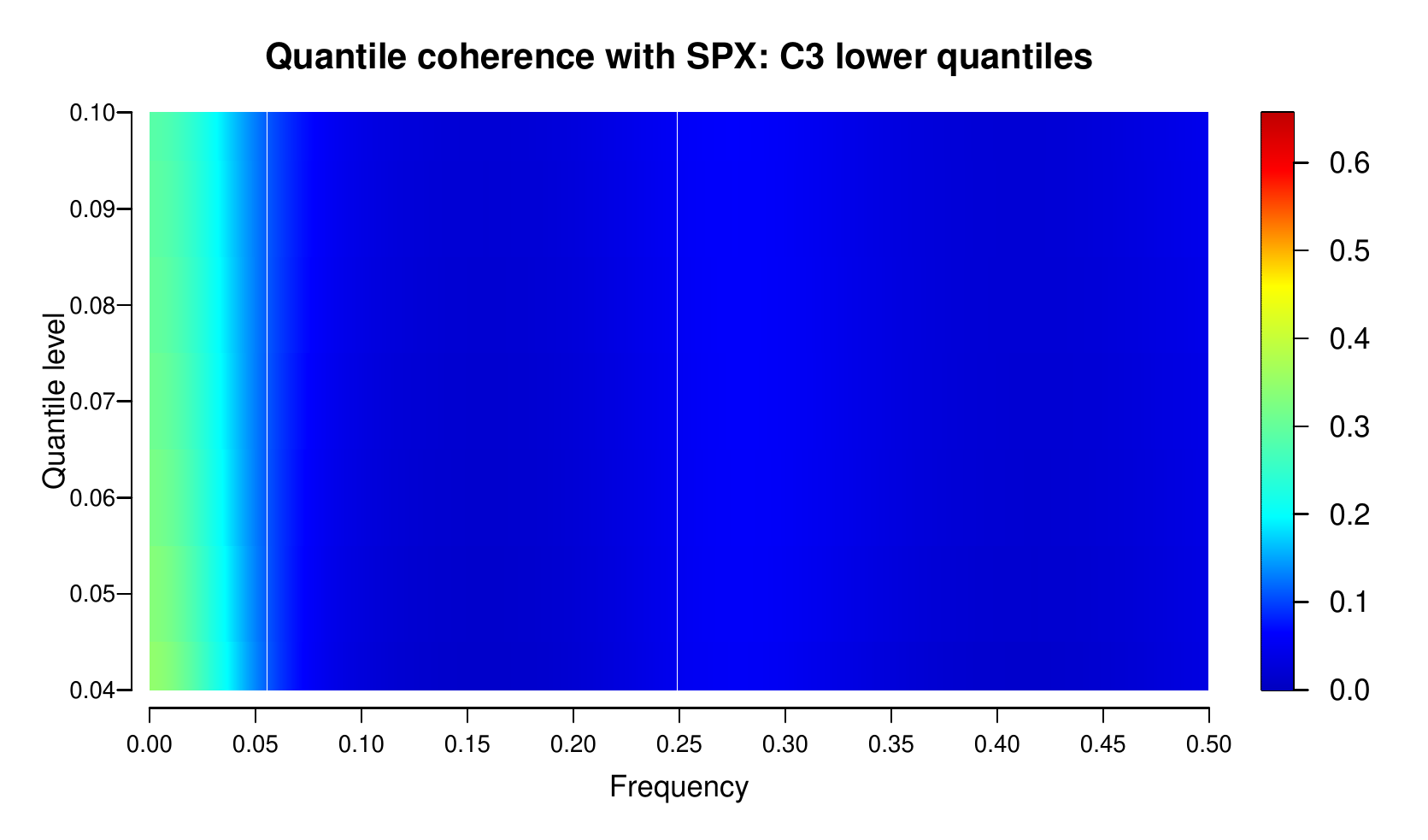}
\caption{The quantile coherence spectra for the centroids in each of the 3 clusters with the SPX index. For upper quantiles (first row). For middle quantiles ( second row). For lower quantiles (third row).}
\label{centroid:quantile_regions}
\end{figure}

To provide a concise summary of the information contained within each of the three clusters within each quantile region, we calculate the centroids of each cluster by averaging their members. The centroids of the three clusters for each specific quantile region are visualized in Figure~\ref{centroid:quantile_regions}. The top panel displays the formation of three clusters based on upper quantiles, the center panel illustrates clusters formed using middle quantiles, and the bottom panel showcases the corresponding clusters formed based on lower quantiles.

It is observed that the overall cluster formation follows a consistent pattern in terms of the distribution of quantile coherence across quantile regions. In particular, Cluster 1 is composed of stocks exhibiting the highest quantile coherence with the SPX. Cluster 2  comprises stocks with intermediate quantile coherence with the SPX, while Cluster 3 consists of stocks with the lowest quantile coherence with the SPX.
It is worth noting that for the clusters formed based on the middle quantile regions, the level of quantile coherence is considerably lower compared to the other quantile regions across all clusters.

Cluster 1, depicted in the left column of Figure~\ref{centroid:quantile_regions}, exhibited high quantile coherence with the SPX index at low frequencies. Interestingly, this cluster also demonstrated significant quantile coherence in the mid-range and high-frequency regions. Notably, among the three clusters, only Cluster 1 displayed such a level of quantile coherence in the middle and high-frequency regions. Further analysis revealed that Cluster 1, based on both lower and upper quantile regions, exhibited the highest level of quantile coherence compared to the middle quantile levels. These findings evidence the importance of quantile dependence in the data.

Cluster 2, illustrated in the center column of Figure~\ref{centroid:quantile_regions}, consisted of stocks that exhibited a moderate level of quantile coherence with the SPX index in the lower frequency regions within both the lower and upper quantile regions. The quantile coherence in Cluster 2 was stronger than that observed in Cluster 3 but not as high as in Cluster 1. Cluster 2 also contained the largest number of stocks among all quantile regions. Similar to Cluster 3, it lacked significant quantile coherence in other frequency ranges; however, the level of quantile coherence in Cluster 2 was notably higher than in Cluster 3.

Finally, in Cluster 3, displayed in the right column of Figure~\ref{centroid:quantile_regions}, stocks in this cluster showed a low level of quantile coherence with the SPX index around low-frequency regions in both the lower and upper quantile regions. However, the quantile coherence observed in Cluster 3 was not as strong as that seen in Clusters 1 and 2. Additionally, no significant quantile coherence activity was observed in other frequency regions. Notably, Cluster 3, based on middle quantiles, demonstrated a very low level of quantile coherence across all frequency regions.

\subsection{Capital Asset Pricing Model (CAPM)}

The beta coefficient in the Capital Asset Pricing Model (CAPM), a time domain approach, is a standard method of quantifying the systematic risk of equities in the financial industry \citep{sharpe1964capital,lintner1965security,mossin1966equilibrium}. The CAPM methodology distinguishes the stocks that are more sensitive to market movements and those that are less sensitive to such changes. Given that no shocks or other factors caused substantial volatility throughout the period studied, the beta coefficients do not have large values. Stocks with betas above 1 will tend to move with more momentum than the S\&P; Stocks with betas less than 1 with less momentum.

The application of the quantile coherence approach reveals clusters that display a certain level of connection with the beta distribution. However, due to their coherence-quantile relationship, the quantile coherence method offers additional insights into their associations with the SPX. As evident from Table \ref{betas_summary}, we can observe two distinct groups of clusters in terms of the beta distribution across all quantile regions. The first group is Cluster 1 which clearly shows the presence mostly of stocks with high betas (higher than 1). The second group is the one formed with Clusters 2 and 3 which represent stocks with betas on average less than 1. Clusters 2 and 3 are difficult to distinguish from the beta distribution since their beta-value distributions overlap. Using the quantile coherence, these two clusters are distinguished based on their activity connected with specific quantile areas (See Figure~\ref{centroid:quantile_regions}).  
\begin{table}[!htb]
\centering
\caption{Summary statistics for the beta values obtained for each of the three clusters obtained through quantile coherence}
\resizebox{0.7\textwidth}{!}{
\begin{tabular}{llllll}
\toprule
\textbf{Quantiles} & \textbf{Clusters}  & \textbf{Min} & \textbf{Max}& \textbf{Mean} & \textbf{Sd} \\
\midrule
\multirow{3}{*}{\textbf{Lower}} & \textbf{Cluster 1} & 0.8563 &1.5826 &1.1079 &0.1907  \\ \cmidrule{2-6}
                                          & \textbf{Cluster 2} & 0.4821 &1.2146 &0.8070 & 0.2368 \\ \cmidrule{2-6}
                                          & \textbf{Cluster 3} &0.4360 &0.7903 &0.5489 & 0.1421  \\  \midrule
\multirow{3}{*}{\textbf{Middle}}          & \textbf{Cluster 1} &0.8563 &1.5826 &1.1420 &0.1890   \\ \cmidrule{2-6}
                                          & \textbf{Cluster 2} &0.5070 &1.2146 &0.8247 &0.2392 \\ \cmidrule{2-6}  
                                          & \textbf{Cluster 3} &0.4360 &0.8427 &0.5779 & 0.1535 \\
                                          \midrule
\multirow{3}{*}{\textbf{Upper}}           & \textbf{Cluster 1} &0.8563 &1.5826 &1.1410 &0.1873 \\ \cmidrule{2-6}
                                          & \textbf{Cluster 2} &0.5194 &1.2146 &0.8524 &0.2179 \\ \cmidrule{2-6}
                                          & \textbf{Cluster 3} & 0.4360 &0.8427 &0.5700 &0.1455  \\ 
\bottomrule
\end{tabular}}
\label{betas_summary}
\end{table}

\subsection{Comparison with ordinary VAR coherence}

\begin{figure}
    \centering
    \includegraphics[width=1\linewidth]{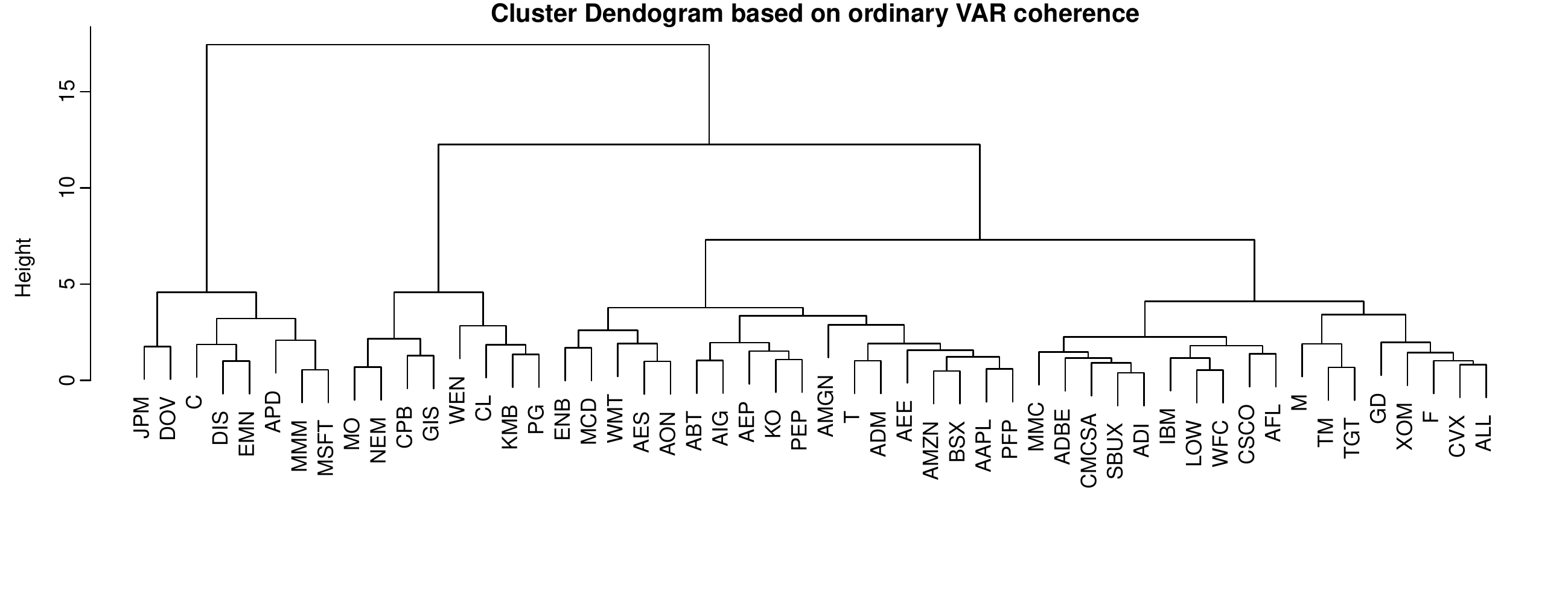}
    \caption{Dendrogram for the hierarchical clustering of  52  stocks in SPX based on the ordinary coherence with the index.}
    \label{app:orddendrogram}
\end{figure}

\begin{figure}
    \centering
    \includegraphics[width=0.7\textwidth]{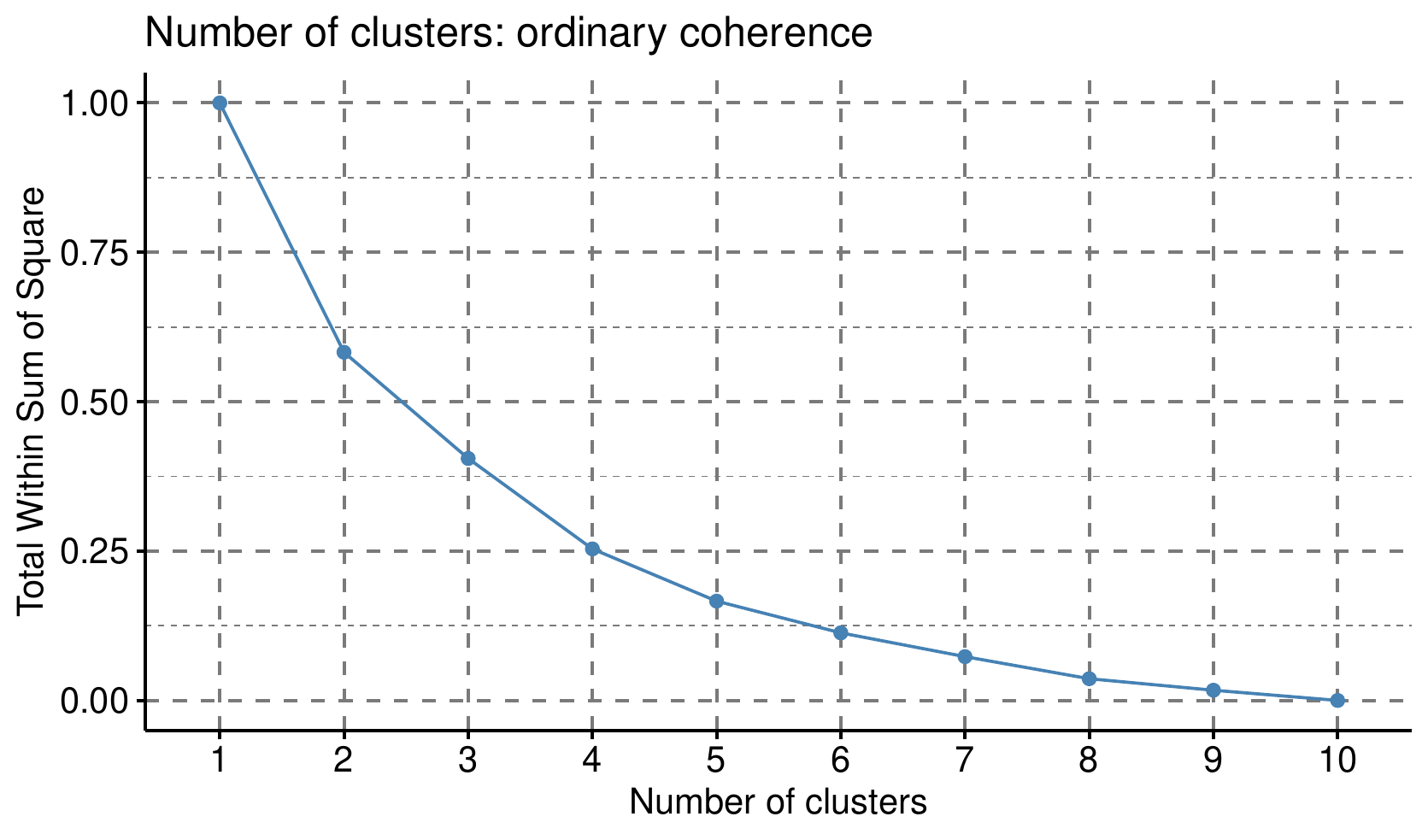}
    \caption{The total within-cluster sum of Squares (WSS) as a function of the number of clusters.}
    \label{app:optimal_ord}
\end{figure}

For comparison, we repeat the hierarchical clustering procedure using the ordinary VAR coherence.  The resulting dendrogram is shown in Figure~\ref{app:orddendrogram}. To select the optimal number of clusters in this case, similar to the quantile coherence case, the elbow plot is in Figure~\ref{app:optimal_ord}. From Figure~\ref{app:optimal_ord} we can see, that determining an optimal number of clusters is difficult because no clear turning point can be identified. With the sole purpose of making a fair comparison between the quantile coherence and the ordinary coherence when doing clustering, we use the same number of clusters for the ordinary coherence.

\begin{figure}[!htb]
\centering
    \includegraphics[width=0.7\linewidth]{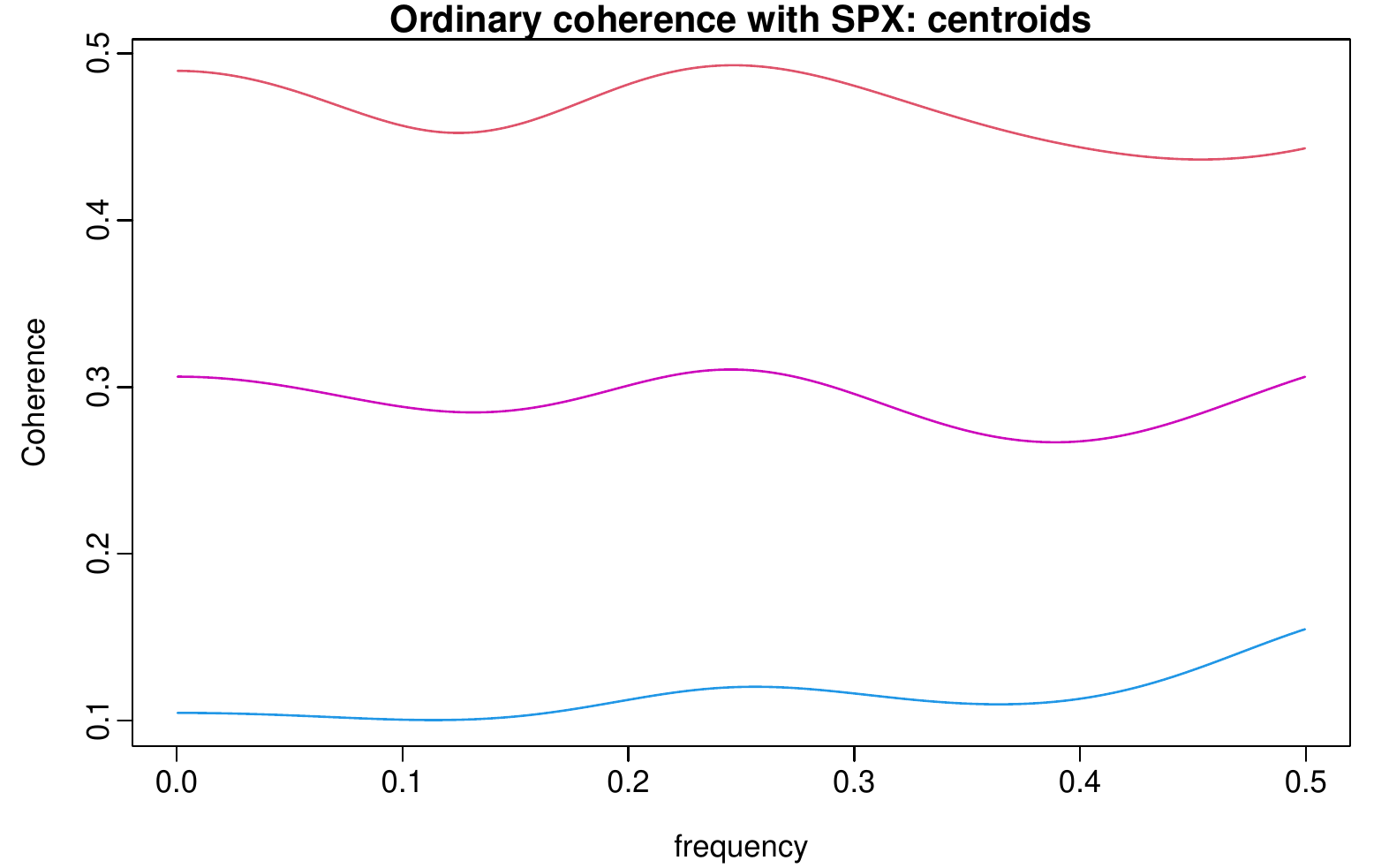}
    \caption{The ordinary coherence spectra for the centroids in each of the 3 clusters with SPX.  }
    \label{app:centroind_oc}
\end{figure}

The analysis of the ordinary coherence reveals distinct behaviors within the three clusters, as depicted by the centroids in Figure~\ref{app:centroind_oc}. Cluster 1 exhibits high coherence, Cluster 2 demonstrates moderate coherence, and Cluster 3 displays very low coherence.  The stock assignment based on ordinary coherence can be found in Table~\ref{Tab:stocks_ordinary_coh}.

\begin{table}[h]
\resizebox{\textwidth}{!}{
\begin{tabular}{ll}
\toprule
 \textbf{Clusters}  &  \textbf{Stock Members}  \\ \midrule
 \textbf{Cluster 1} & APD C DIS DOV EMN JPM MMM MSFT     \\ \midrule
                \multirow{3}{*}{\textbf{Cluster 2}} &   AAPL ADBE AEE AEP AES AFL AIG ALL AMGN AMZN  APD BSX C    \\ 
                                     &  CVX CSCO  DIS DOV EMN ENB F GD IBM KO LOW M  MCD MMC    \\
                                     &  MMM MSFT PEP PFP SBUX T TGT TM WFC WMT XOM\\  \midrule
                                     
                                     \textbf{Cluster 3}   &  CL CPB GIS KMB MO NEM PG WEN      \\
                                          \bottomrule
\end{tabular}}
\caption{Stock assignment in hierarchical clustering based on ordinary coherence}
\label{Tab:stocks_ordinary_coh}
\end{table}

The main differences compared to quantile coherence can be observed in Cluster 2, which contains the largest number of stocks. Cluster 1, characterized by quantile coherence, is a more compact cluster that enables a clear distinction between high and moderate quantile coherence, particularly at higher and lower quantile regions. In contrast, when considering ordinary coherence, many stocks initially assigned to Cluster 1 by quantile coherence now belong to Cluster 2.

By examining the centroid of Cluster 2 using quantile coherence, specifically at higher quantile regions (refer to Figure \ref{centroid:quantile_regions}), it becomes apparent that these stocks are expected to exhibit substantial coherence around the zero frequency region, similar to Cluster 1. Additionally, there is observable activity in the middle and higher frequency regions. However, the level of quantile coherence in the zero-frequency zone is not as pronounced as that of Cluster 1. 

\begin{figure}[h]
\centering
\includegraphics[width=0.45\textwidth]{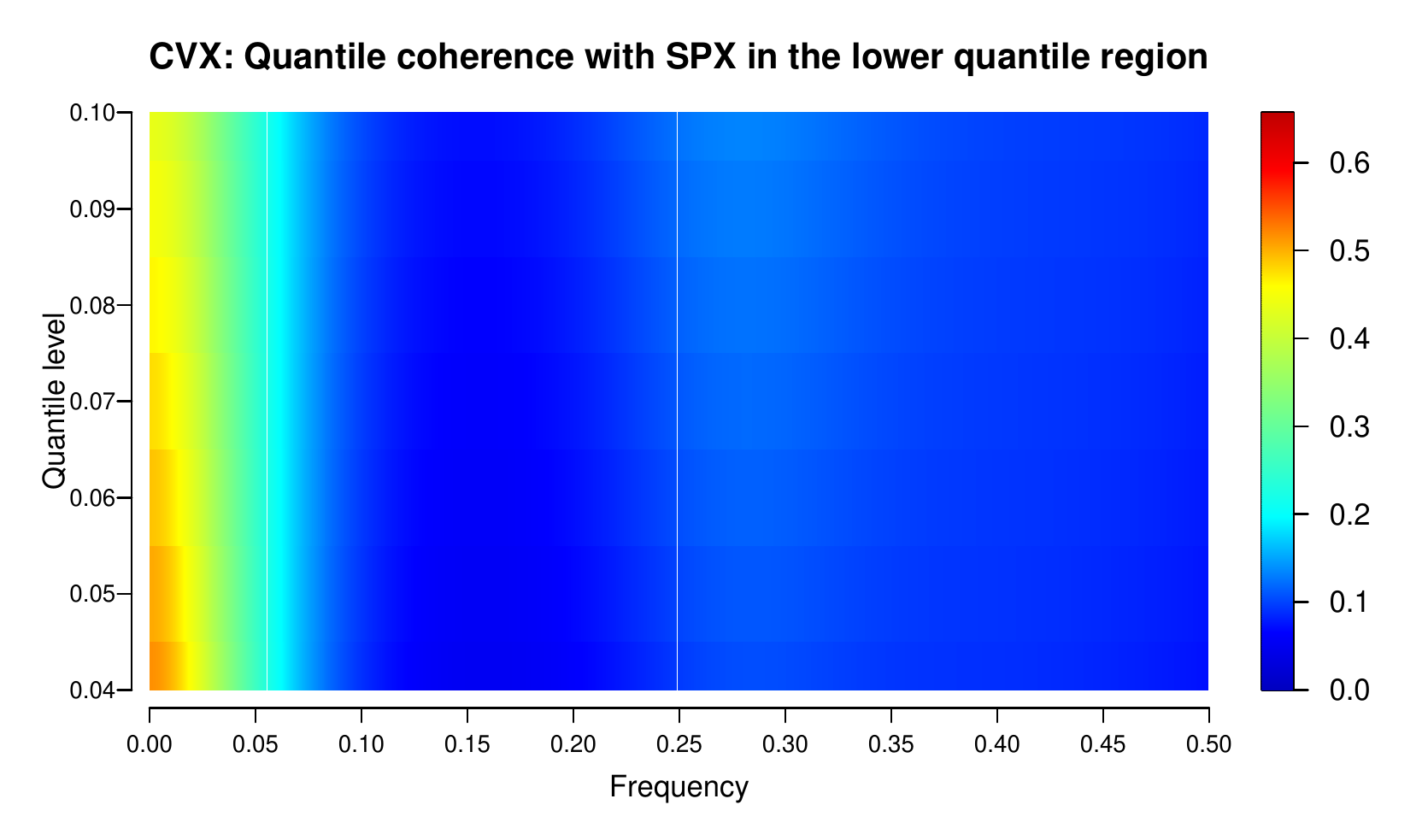}
\qquad
\includegraphics[width=0.45\textwidth]{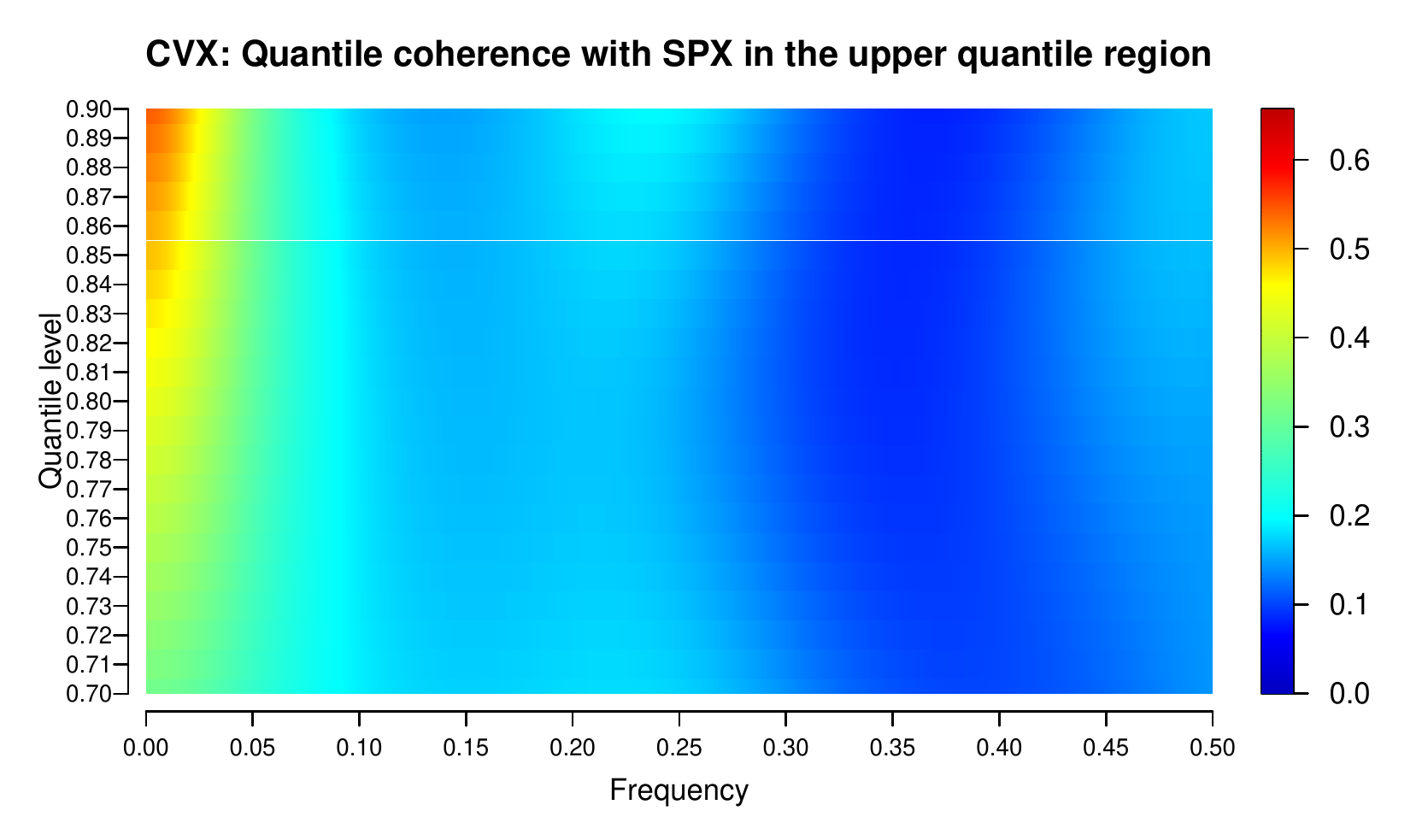}
\qquad
\includegraphics[width=0.45\textwidth]{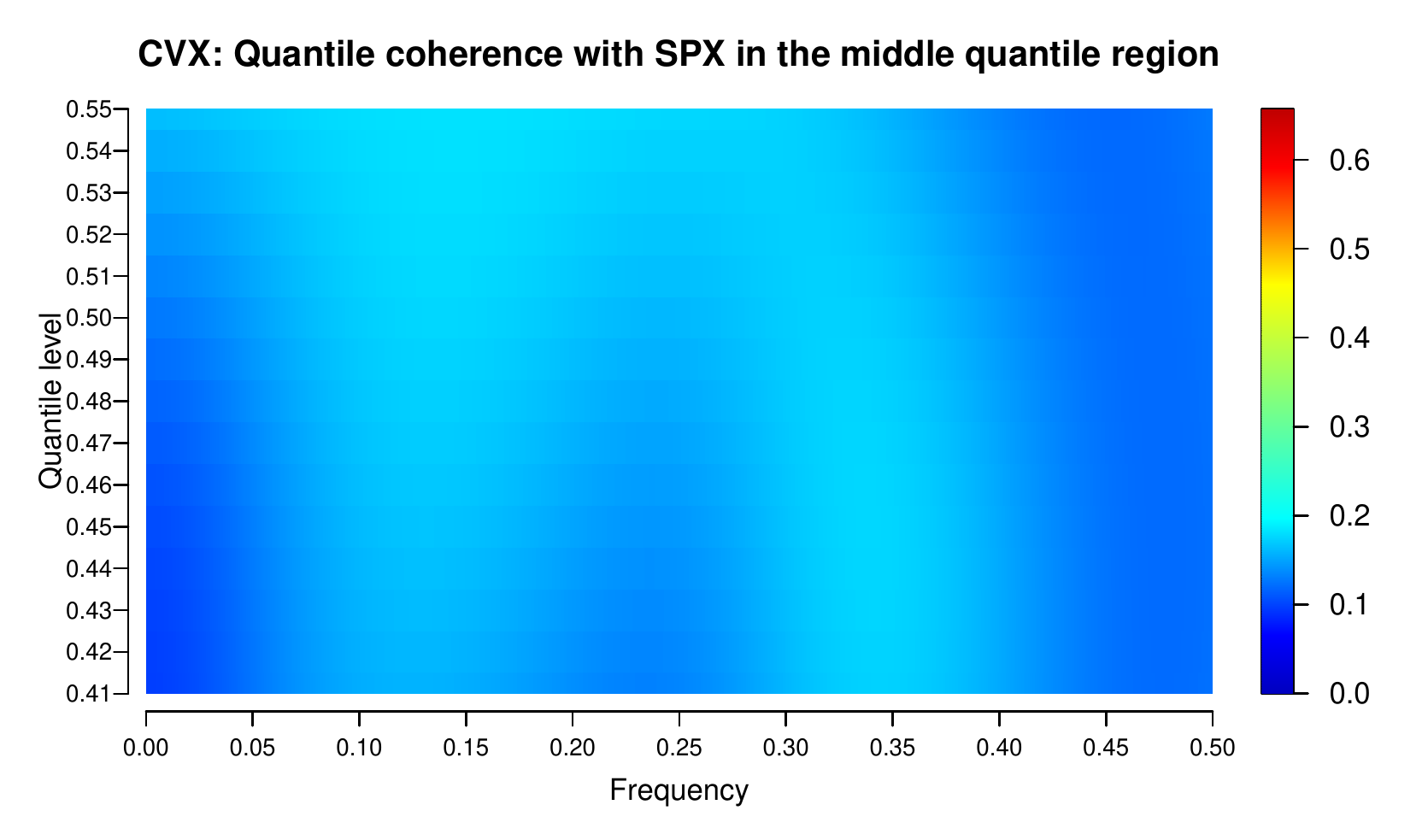} 
\caption{quantile coherence with SPX for the stock CVX. Top left panel lower quantiles. Top right panel upper quantiles. Bottom panel middle quantiles.}
\label{CVX}
\end{figure}

Finally, we have chosen two examples to illustrate the enhanced characteristics of hierarchical clustering when utilizing quantile coherence at specific quantile regions. Furthermore, we have also included an example that compares quantile coherence and ordinary coherence when clustering. In Figure~\ref{CVX}, we provide an analysis of the quantile coherence for the stock "CVX" in relation to the SPX, focusing on different quantile regions. The top panel displays the quantile coherence at lower and upper quantile levels, while the bottom panel represents the coherence at the middle quantile level.

When clustering "CVX" based on both upper and lower quantile levels, it is assigned to cluster 2, which corresponds to a cluster exhibiting a moderate or middle level of quantile coherence. However, when clustering is performed using the middle quantile levels, "CVX" is assigned to cluster 1. According to the general behavior, stocks in cluster 1 are expected to demonstrate the highest level of quantile coherence, particularly around the zero frequency region. However, in the case of "CVX," we observe that at the middle quantile levels, it does not exhibit a sufficiently high level of quantile coherence to be assigned to Cluster 1.

\begin{figure}[h]
\centering
\includegraphics[width=0.4\textwidth]{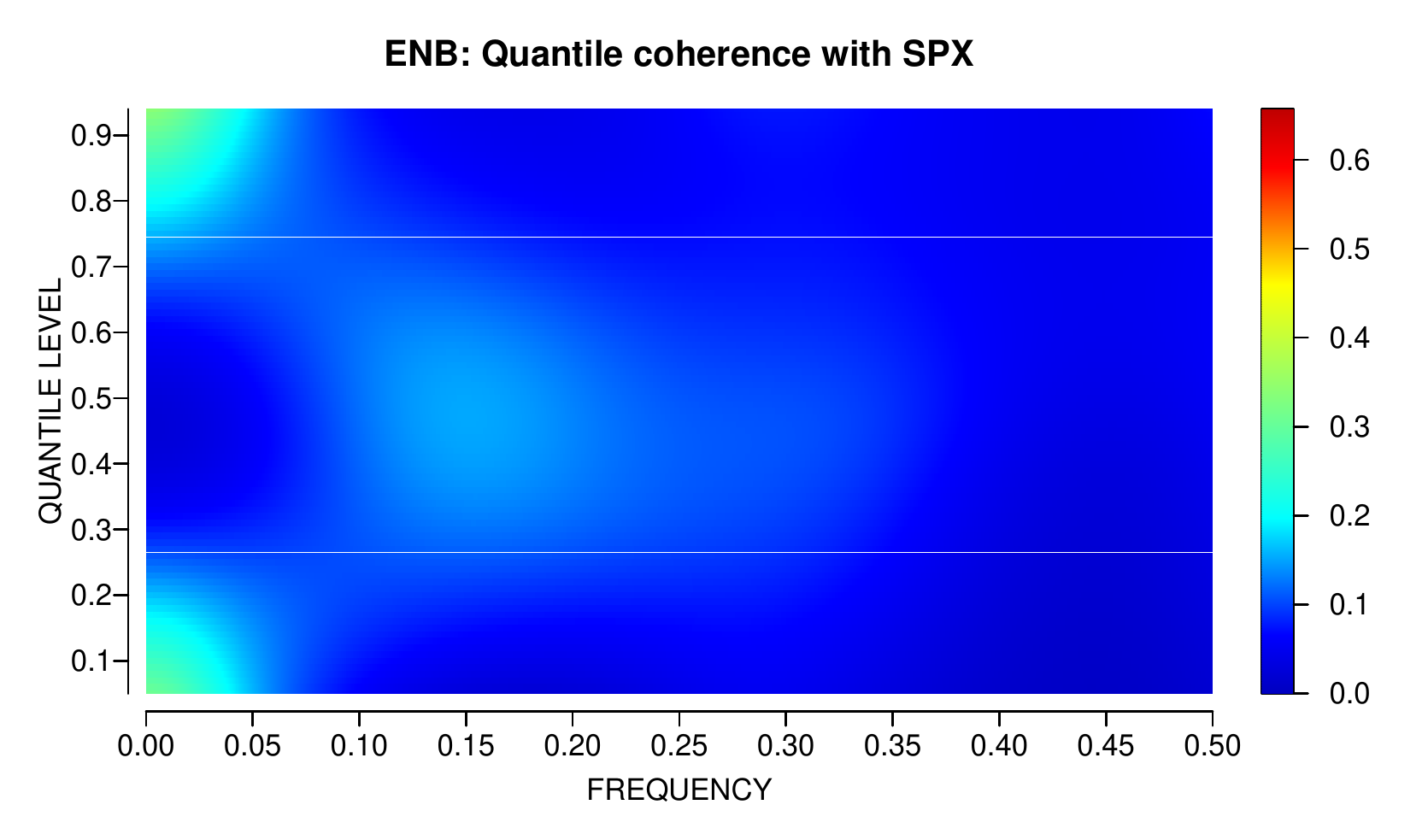}
\qquad
\includegraphics[width=0.4\textwidth]{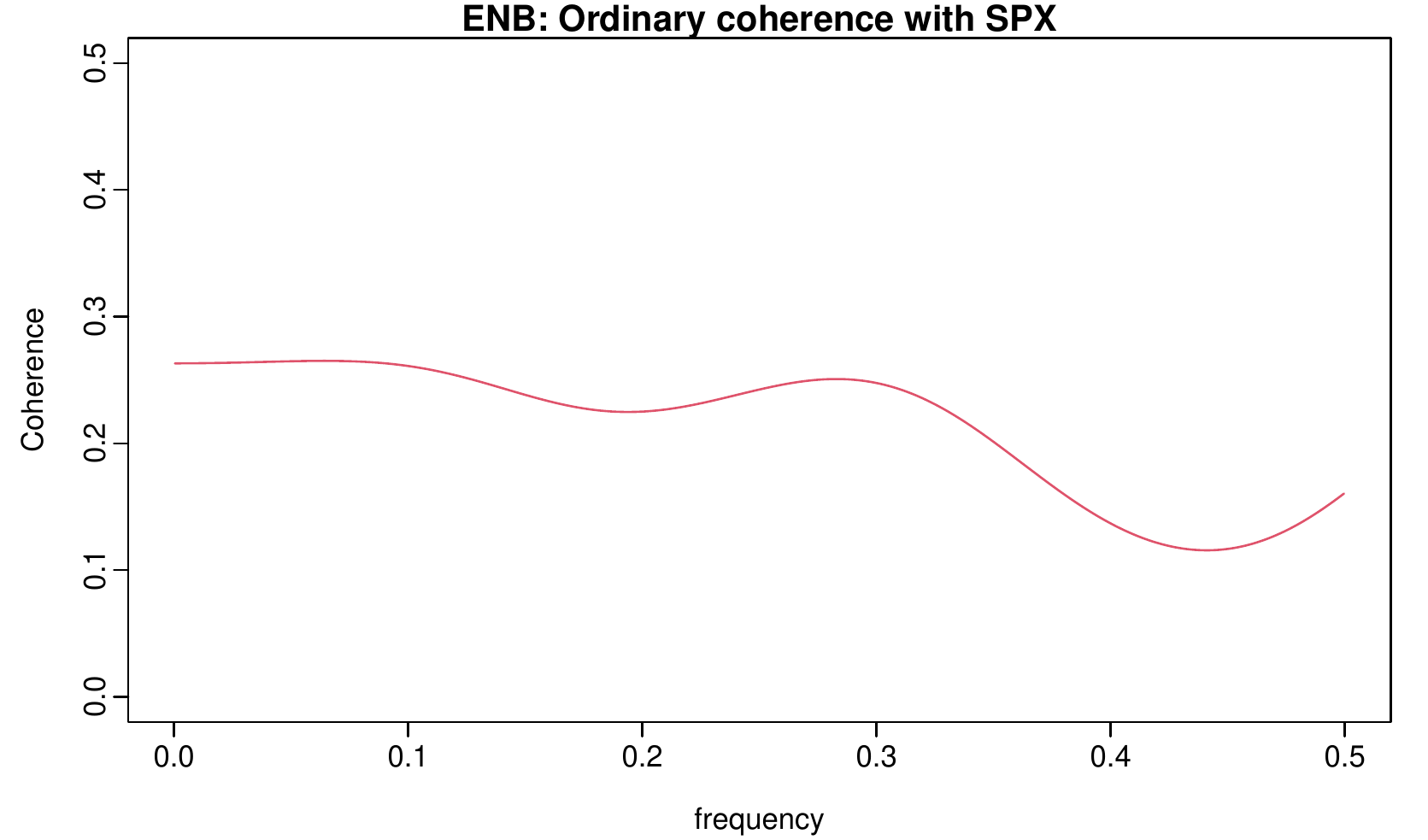}
\caption{ Left panel: quantile coherence with SPX for the stock ENB.  Right panel: ordinary coherence with SPX for the stock ENB. }
\label{ENB}
\end{figure}

Figure~\ref{ENB} illustrates the comparison between quantile coherence (left panel) and ordinary coherence (right panel) for the stock "ENB" in relation to SPX. It is evident that "ENB" exhibits generally low values of quantile coherence across all quantile levels. Consequently, when considering each of the three quantile regions, "ENB" is assigned to Cluster 3, which is characterized by a low level of quantile coherence. However, when examining the ordinary coherence, "ENB" demonstrates a moderate level of coherence. As a result, it is assigned to Cluster 2 based on the ordinary coherence criterion as it does not consider the quantile information.

\section{Discussion}\label{sec:discussion}

We developed a new semi-parametric method for estimating the quantile coherence derived from trigonometric quantile regression. The proposed method employs the parametric form of the VAR approximation in conjunction with a nonparametric smoothing technique. The parametric VAR spectrum estimates the multivariate quantile spectral matrix at each quantile. The VAR model is obtained by solving the multivariate Yule-Walker equations formed by the quantile autocovariance function which is defined as the inverse Fourier transform of quantile periodograms. The AIC criterion, which balances the goodness of fit and the model complexity is used to determine the VAR order. The resulting preliminary estimate of the quantile coherence is further smoothed across quantiles by smoothing splines where the smoothing parameter is selected jointly across frequencies. For selecting the tuning parameter, a $\mathcal{K}$-fold cross-validation technique is employed to cope with the correlation found in the estimated quantile coherence across quantiles.

Similarly to the quantile periodogram maps previously described in the literature \citep{Li2012, Li2020}, the 2D representation of the quantile coherence provides more information than the ordinary coherence and can be used as images to analyze multivariate time series data. 

We also presented the result of an application of quantile coherence to financial time series. In this application, the daily closing prices of 52 stocks are grouped by their behavior against the SPX index as measured by the quantile coherence. The three clusters are distinguished largely by the coherence patterns in the low-frequency region at high and/or low quantiles. A better cluster formation is obtained by focusing on specific quantile regions. The quantile-dependent nature enhances cluster formation.

\bibliographystyle{chicago}
\bibliography{paper2}
\clearpage
\appendix

\section{Appendix}

\subsection{Multivariate Durbin-Levinson Algorithm and the standard partial autocorrelation function}\label{methods: Multivariate Levinson D algo}

\cite{Whittle1963} proposed the multivariate version of the Durbin-Levinson recursions, which solve the Yule-Walker equations in \eqref{YW2} with a total of $2p$ inversions of $k\times k$ matrices. \citep[see also proposition 11.4.1][]{Brockwell1992}. 

\cite{degerine1990} proposed a way of defining the partial autocorrelation function for multivariate stationary time series through the canonical analysis of the forward and backward innovations. It is proved that there is a one-to-one correspondence between the resulting PACF and the autocovariance function. 

Based on \cite{Whittle1963} and the canonical approach defined by \cite{degerine1990}, we employ the following algorithm to compute the VAR parameters recursively. Given a sequences of covariance matrices ${\bm{\Gamma}(0), \bm{\Gamma}(1), \ldots \bm{\Gamma}(p) }$, 
\begin{enumerate}[noitemsep]
\setcounter{enumi}{-1}
    \item Compute $ V_0= \widetilde{V_0}=\bm{\Gamma}(0)$ and $\Delta_0 = \bm{\Gamma}(1)$.
\item For $r=1,\ldots,p$,  compute the PACF $\mathbf{\Psi}_r=\mathbf{V}^{-1/2}_{r-1}\Delta_{r-1} \widetilde{\mathbf{V}}_{r-1}^{-1/2}$.
\item Compute 
\vspace{-1cm}
\begin{align*}
    \mathbf{\Phi}_{r,r}         &= \mathbf{V}^{1/2}_{r-1} \mathbf{\Psi}_r \widetilde{\mathbf{V}}_{r-1}^2\\
    \widetilde{\mathbf{\Phi}}_{r,r} &= \widetilde{\mathbf{V}}_{r-1}^{1/2} \mathbf{\Psi}_r^\top \mathbf{V}^{1/2}_{r-1}.
\end{align*}
\item For $p \geq 2$, compute
\begin{align*}
    \mathbf{\Phi}_{r,r'}         &= \mathbf{\Phi}_{r-1,r'}-\mathbf{\Phi}_{r,r}\widetilde{\mathbf{\Phi}}_{r-1,r-r'},~r'=1,\ldots r-1 \\
    \widetilde{\mathbf{\Phi}}_{r,r'} &= \widetilde{\mathbf{\Phi}}_{r-1,r'}-\widetilde{\mathbf{\Phi}}_{r,r}\mathbf{\Phi}_{r-1,r-r'},~r'=1,\ldots r-1.
\end{align*}
\item Compute
\begin{align*}
    \mathbf{V}_r &= \mathbf{V}^{1/2}_{r-1}[\mathds{I}-\mathbf{\Psi}_r\mathbf{\Psi}^\top_r]\mathbf{V}^{1/2\top}_{r-1}\\
             \widetilde{\mathbf{V}}_{r} &= \widetilde{\mathbf{V}}^{1/2}_{r-1}[\mathds{I}-\mathbf{\Psi}^\top_r\mathbf{\Psi}_r]\widetilde{\mathbf{V}}^{1/2\top}_{r-1}\\
    \mathbf{\Delta}_r &= \bm{\Gamma}(r+1)-  \mathbf{\Phi}_{r,1}\bm{\Gamma}(r)-\ldots-\mathbf{\Phi}_{r,r}\bm{\Gamma}(1).
\end{align*}

\end{enumerate}

For the VAR model to be stable, the singular values of the PACF $(\Psi_r)$ must be less than  $1$ in magnitude.  \cite{Morf1978} proposed the orthogonalization procedure based on a Gramm-Schmidt process to compute the PACF and the resulting Durbin-Levinson recursion. 

\subsection{Proof of Equation~\eqref{ord_acf}}\label{A3:proof}

By the $2n$-point inverse discrete Fourier transform, the estimated autocovariance function $\widehat{\mathbf{\Gamma}}(h): |h|<n$ can be recovered from Equation~\eqref{ord_per}, as follows:

Multiplying both sides of Equation~\eqref{ord_per} by $e^{i2\pi l k/2n}$ and summing over $l: 0 \leq l \leq 2n-1$, 
\begin{align*}
    (2n)^{-1}\sum_{l=0}^{2n-1} \mathbf{I}_n(\omega_l)e^{i 2\pi l k/2n}&=(2n)^{-1}\sum_{l=0}^{2n-1} \sum_{|h|<n} \widehat{\mathbf{\Gamma}}(h) e^{-i 2 \pi l k/2n}e^{i2\pi l h/2n}\\
    &=(2n)^{-1}\sum_{|h|<n} \widehat{\mathbf{\Gamma}}(h)\left[\sum_{l=0}^{2n-1} e^{i 2 \pi (h-k) l/2n}\right],\\
    &\text{with}~h=k,\\
    &=\widehat{\mathbf{\Gamma}}(h),h=0,1,\cdots,n-1\\ 
\end{align*}

\clearpage

\subsection{Hierarchical clustering based on specific quantile regions}\label{A2:dend}

\begin{figure}[h]
\centering
\includegraphics[width=1\linewidth,height=4.9 cm]{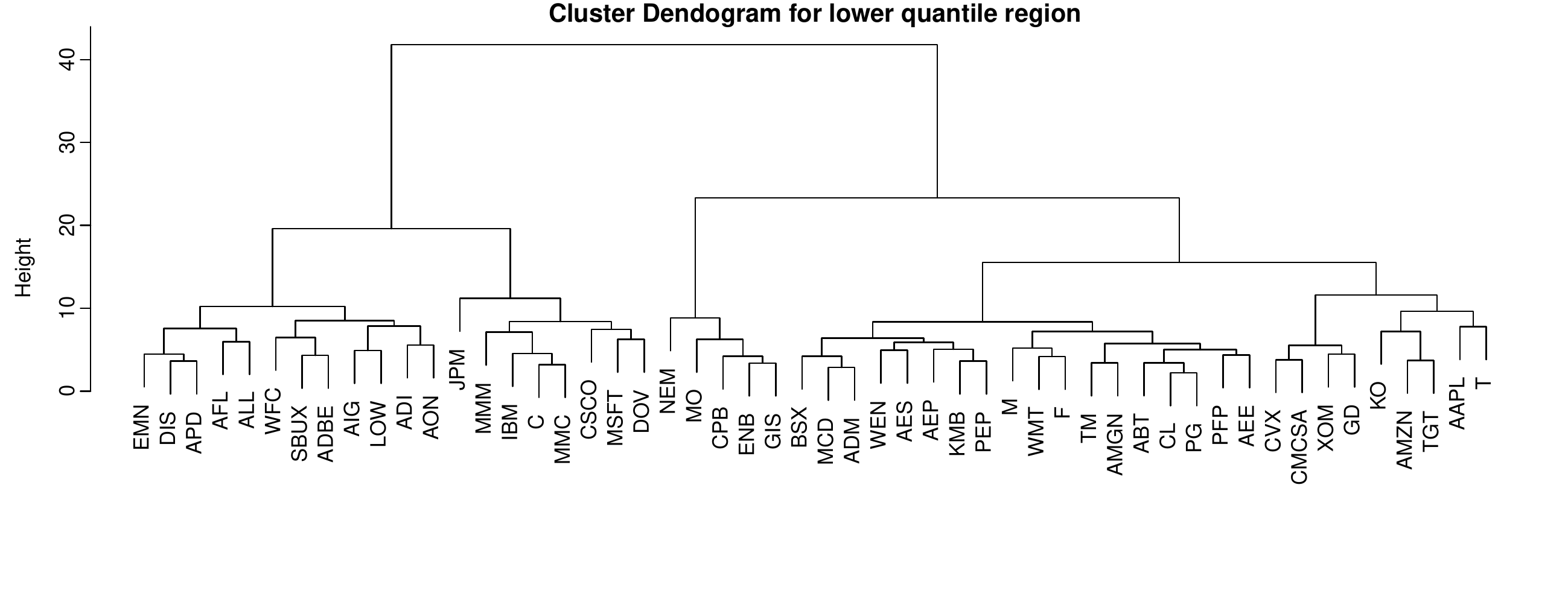}
\\
\includegraphics[width=1\linewidth,height=4.9 cm]{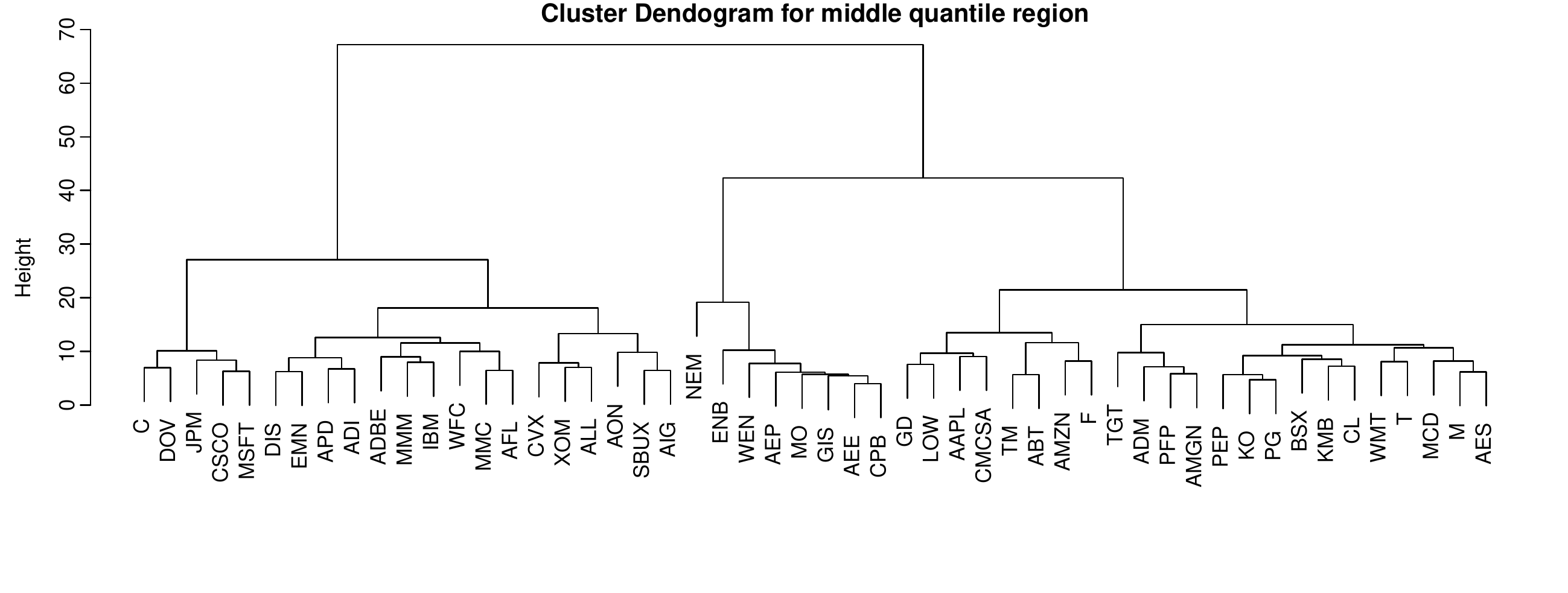}
\\
\includegraphics[width=1\linewidth,height=4.9 cm]{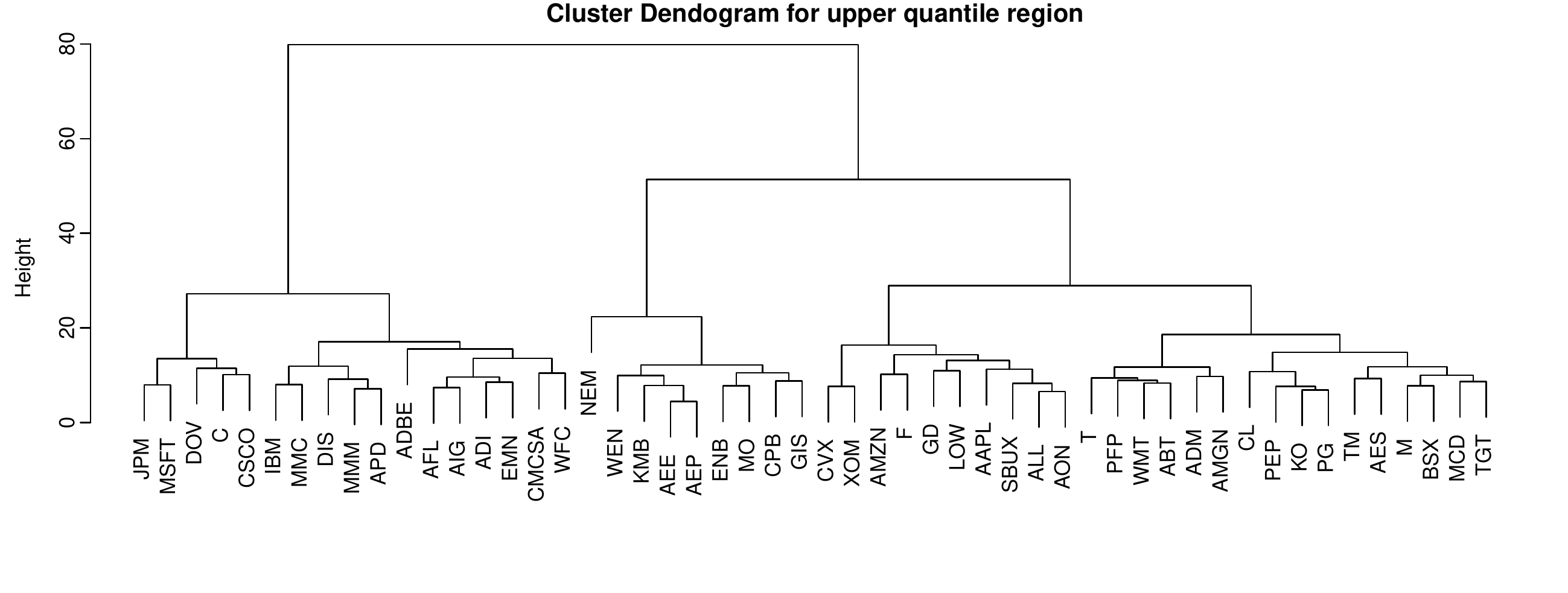}
\caption{Dendrogram for the hierarchical clustering of 52 stocks in SPX based on the quantile coherence of their daily log returns with the index SPX. The top panel displays clusters formed using quantile coherence for lower quantile levels. The center panel depicts clusters based on quantile coherence for middle quantile levels. The bottom panel represents clusters based on quantile coherence for upper quantile levels.}
\label{Application:dend_upper_lower}
\end{figure}

\end{document}